\documentstyle[12pt,fleqn]{article}
\def\thesection{}

\def\msun{M_{\odot}}
\def\be{\begin{equation}}
\def\ee{\end{equation}}
\def\bea{\begin{eqnarray}}
\def\eea{\end{eqnarray}}
\def\re#1{{[\ref{#1}]}}
\def\kay{{{\bf \vec{k}}}}
\def\exx{{{\bf \vec{x}}}}
\def\la{\mathrel{\mathpalette\fun <}}

\def\fun#1#2{\lower3.6pt\vbox{\baselineskip0pt\lineskip.9pt
        \ialign{$\mathsurround=0pt#1\hfill##\hfil$\crcr#2\crcr\sim\crcr}}}
\begin{document}
\begin{titlepage}
\vspace*{-62pt}
\begin{flushright}
{\small
SUSSEX-AST 93/3-1 \\
FNAL--PUB--93/029-A\\
March 1993}
\end{flushright}
\begin{center}
{\Large \bf Reconstructing the inflaton potential---\\ in principle and in
practice}\\
\vspace{0.6cm}
\normalsize
Edmund J.\ Copeland$^1$\\
{\em School of Mathematical and Physical Sciences,\\
University of Sussex, Brighton BN1 9QH, U.\ K.}\\
\vspace{0.4cm}
Edward W.\ Kolb$^2$\\
{\em NASA/Fermilab Astrophysics Center\\
Fermi National Accelerator Laboratory, Batavia, IL~~60510, and\\
Department of Astronomy and Astrophysics, Enrico Fermi Institute\\
The University of Chicago, Chicago, IL~~ 60637}\\
\vspace{0.4 cm}
Andrew R.\ Liddle$^3$\\
{\em Astronomy Centre, School of Mathematical and Physical Sciences,\\
University of Sussex, Brighton BN1 9QH, U.\ K.}\\
\vspace{0.4cm}
James E.\ Lidsey$^4$\\
{\em Astronomy Unit, School of
Mathematical Sciences,\\
Queen Mary and Westfield College, Mile End Road,
London E1 4NS, U.\ K., and\\
NASA/Fermilab Astrophysics Center\\
Fermi National Accelerator Laboratory, Batavia, IL~~60510}
\end{center}

\vspace*{12pt}

\begin{quote}
\normalsize
\hspace*{2em} Generalizing the original work by Hodges and Blumenthal, we
outline a formalism which allows one, {\em in principle,} to reconstruct the
potential of the inflaton field from knowledge of the tensor gravitational
wave spectrum or the scalar density fluctuation spectrum, with special
emphasis on the importance of the tensor spectrum.  We provide some
illustrative examples of such reconstruction. We then discuss in some detail
the question of whether one can use real observations to carry out this
procedure. We conclude that {\em in practice,} a full reconstruction of the
functional form of the potential will not be possible within the foreseeable
future.  However, with a knowledge of the dark matter components, it should
soon be possible to combine intermediate-scale data with measurements of
large-scale cosmic microwave background anisotropies to yield useful
information regarding the potential.

\vspace*{12pt}

PACS number(s): 98.80.--k, 98.80.Cq, 12.10.Dm

\noindent
\small email: $^1$edmundjc@central.sussex.ac.uk;~~~$^2$rocky@fnas01.fnal.gov;\\
\phantom{email:} $^3$arl@starlink.sussex.ac.uk;~~~$^4$jim@fnas09.fnal.gov

\end{quote}

\normalsize

\end{titlepage}

\thesection{\centerline{\bf I. INTRODUCTION}}
\setcounter{section}{1}
\setcounter{equation}{0}
\vspace{18pt}
The detection by the {\em COBE} DMR instrument of fluctuations in the
temperature distribution of the Cosmic Microwave Background Radiation (CMBR)
on large angular scales \re{DMR} is certainly one of the most significant
cosmological results since the detection of the CMBR itself. These
fluctuations provide valuable information about the nature of primordial
perturbations believed responsible for the origin of structure in the
Universe.  The horizon radius at the epoch of last scattering of the CMBR
corresponds to angular scales of about $2^\circ$ on the sky, which implies
that fluctuations on scales probed by {\em COBE} were not predominantly
affected by causal processes or the nature of the matter constituents of the
Universe at the time of last scattering of the CMBR. Indeed, the large-scale
(greater than $2^\circ$) fluctuations arise from the Sachs--Wolfe effect when
photons are either red or blue shifted as they climb out of, or fall into,
gravitational potential wells \re{SW}. It is most likely that the fluctuations
in the CMBR are the result of processes that occurred very early in the history
of the Universe, so they yield vital information concerning the physics that
led to the primordial perturbations.

There are currently two very attractive scenarios for the origin of the
primordial fluctuations: quantum effects during inflation, and gravitational
effects of defects resulting from cosmological phase transitions.  Both
scenarios involve physics beyond the standard model of particle physics,
involving energies in the range $10^{10}$GeV $\la E \la 10^{19}$GeV, an energy
scale we will refer to loosely as the Grand Unified Theory (GUT) scale.  A
major difference in the predictions of the two scenarios concern the Gaussian
nature of the fluctuation pattern, and we should be able to use this to
differentiate between the two possibilities in the near future.  In this paper
we will assume that the fluctuations are the result of inflation, and we
discuss what might be learned about particle physics at very high energies from
astronomical observations from which we can infer the primordial fluctuation
spectrum.

All models of inflation involve a period of rapid growth of the size of the
Universe.  This is most easily illustrated by considering a homogeneous,
isotropic Universe with a flat Friedmann--Robertson--Walker (FRW) metric
described in term of a scale factor $a(t)$.  Here, ``rapid growth'' means a
positive value of $\ddot{a}/a=-(4\pi G_N/3) (\rho+3p)$ where $\rho$ is the
energy density and $p$ the pressure.  In all successful models of inflation,
the Universe is dominated by some sort of scalar ``potential'' energy density
$V>0$ that is positive, resulting in an effective equation of state
$\rho\simeq -p\simeq V$, and hence $\ddot{a}>0$. If one identifies the
potential energy as arising from the potential of some scalar field $\phi$,
then $\phi$ is known as the {\em inflaton} field.

Even within this traditional view of inflation, there are two major ways to
implement the scenario. One way involves a first-order phase transition. In
this method, either in the original proposal of Guth \re{GUTH} or the latest
version called extended inflation \re{FOI},  the inflaton is trapped in a
meta-stable, or false-vacuum, state while the Universe inflates. Inflation is
ended when the Universe undergoes a first-order phase transition in which the
inflaton field tunnels to its true-vacuum state.  In the second method,
inflation occurs because for some reason the inflaton field is displaced from
its minimum and its potential energy density dominates the Universe; inflation
occurs while the inflaton field is slowly evolving, or rolling, to its minimum
[\ref{NEW}, \ref{CHAOTIC}].  It is this second class of ``slow-roll'' models we
will consider in this paper.\footnote{In reality, often the distinction between
the two methods is not so clean, and it is possible to consider some types of
first-order inflation models as variants of slow-roll models. See Ref.\
\re{FOI}.}

Although the early slow-roll models had potentials that were reasonably simple
(Cole\-man--Weinberg, $\lambda\phi^4$, etc.), or at least polynomials in some
scalar field, many attractive models have been developed where the scalar
potential driving inflation is quite complicated.   Perhaps the study of the
density perturbations produced by inflation can shed some light on the nature
of the potential.

Broadly speaking, inflation predicts a very nearly Gaussian spectrum of
density perturbations that is {\em scale dependent}, i.e., the amplitude of
the perturbation depends upon the length scale.  Such a dependence typically
arises because the Hubble expansion rate during the inflationary epoch in fact
changes, albeit slowly, as the field driving the expansion rolls towards the
minimum of the scalar potential. This implies that the amplitude of the
fluctuations as they cross the Hubble radius will be weakly time-dependent.

Within the context of slow-roll inflation, Hodges and Blumenthal \re{HB} have
shown that any scale dependence for density perturbations is possible if one
considers an arbitrary functional form for the inflaton potential, $V(\phi)$.
In this sense, inflation makes no unique prediction concerning the form of the
spectrum and one is left with two options. Either one can aim to find a deeper
physical principle that uniquely determines the potential, or observations that
depend on $V(\phi)$ can be employed to limit the number of possibilities.

Improved observations of large-scale structure, of which {\em COBE} provides
the most dramatic example at present, are important because they allow us, in
principle, to determine the spectrum of primordial density perturbations. This
may very well provide a direct experimental window on the physics of the Grand
Unified era corresponding to energy scales of the order $10^{16}$ GeV.  The
purpose of the present work is to investigate to what extent information from
the CMBR and large-scale galactic structure will allow us to reconstruct  GUT
physics.

In the following section we will review the salient aspects of slow-roll
inflation.  In Section III we discuss the reconstruction of the inflaton
potential from knowledge of scalar or tensor perturbations.  Section IV
illustrates the formalism by several examples in which the functional form of
the potential is found from knowledge of the tensor and scalar perturbation
spectra.  Section V illustrates what can be learned about the potential from
observations of the properties of the tensor and scalar spectra at a
particular length scale.  In Section VI the reader may find a discussion of
how one determines the primordial density spectrum. Finally, Section VII
offers an assessment of the prospectus for reconstruction of the inflaton
potential.

\vspace{48pt}
\thesection{\centerline{\bf II. REVIEW OF SLOW-ROLL INFLATION}}
\setcounter{section}{2}
\setcounter{equation}{0}
\vspace{18pt}

For the benefit of those not familiar with the generation of scalar and tensor
perturbations in slow-roll inflation, we review the salient features in this
section.  Those comfortable with the basic results may wish to skip this
section, and refer back to it as needed to understand notation and
conventions.  We set $c = \hbar = 1$, and define $\kappa^2 = 8\pi G_N =
8\pi/m_{Pl}^2$.

Slow-roll inflation requires a scalar field $\phi$ to be displaced from the
minimum of its potential at some time early in the evolution of the Universe.
If during the evolution of the field to its minimum a region of the Universe
is dominated by the potential energy of the field, then the volume of that
region will undergo rapid expansion, inflate, and grow to encompass a volume
large enough to contain all of the presently observed Universe.  Eventually
the potential energy ceases to dominate when the field evolves through a
steep region of the potential and the field evolves so rapidly that the
kinetic energy of the field comes to dominate.  This is the end of inflation,
and is followed by the scalar field oscillating about the minimum of its
potential, with the inflaton field decaying and `re-heating' the Universe
by conversion of vacuum energy to radiation.

We are interested in the perturbations resulting from inflation. The
``density'' perturbations are usually described in term of fluctuations in the
local value of the mass density. In a Universe with density field
$\rho({\bf x})$ and mean mass density $\rho_0$, the density contrast is defined
as
\be
\label{eq:TWOONE}
\delta ({\bf x})=\frac{\delta \rho({\bf x} )}{\rho_0}
=\frac{\rho({\bf x})-\rho_0}{\rho_0}.
\ee
It is convenient to express this contrast in terms of a Fourier expansion:
\be
\label{eq:FT}
\delta ({\bf x} )  =A  \int \delta_{{\bf k}} \exp(-i{\bf
	k}\cdot{\bf x})d^3\! k,
\ee
where $A$ is simply some overall normalization constant, interesting only for
those who enjoy keeping track of factors of $2\pi$. What is usually meant by
the density perturbation on a scale $\lambda$, $(\delta\rho/\rho)_\lambda$, is
related to the square of the Fourier coefficients $\delta_{{\bf k}}$:
\be
\label{eq:DHROL}
\left(\frac{\delta \rho}{\rho}\right)^2_\lambda \equiv \left. A'
{k^3|\delta_k|^2\over 2\pi^2}\right|_{\lambda=k^{-1}},
\ee
where again we have included an overall normalization constant $A'$.  The
perturbations are normally taken to be (statistically) isotropic, in the sense
that the expectation of $|\delta_{{\bf k}}|^2$ averaged over a large number of
independent regions can depend only on $k =|{\bf k}|$. The dependence of
$\delta\rho/\rho$ as a function of $\lambda$ is the spectrum of the density
perturbations.   Of crucial importance is the relative size of scale $\lambda$
to the scale of the Hubble radius.   The {\em physical} length between two
points of coordinate separation $d$ is $\lambda(t)=a(t)d$.  A length scale
comoving with the expansion will grow proportional to $a(t)$. If
$\ddot{a}(t)<0$, as in the standard non-inflationary phase, then $a(t)$ will
grow slower than $t$.  If $\ddot{a}(t)>0$, as in the inflationary phase, then
$a(t)$ will grow faster than $t$.

For a spatially flat isotropic Universe the Hubble expansion rate, $H(t) =
\dot{a}/a$, is given by $3 H^2(t)=\kappa^2 \rho(t)$.  The inverse of the
Hubble expansion rate, the Hubble radius $\lambda_H(t)\equiv H^{-1}(t)$, is
the scale beyond which causal processes no longer operate.  In the
non-inflationary phase $\lambda_H$ increases linearly with time. Since in the
non-inflationary phase $\lambda_H\propto t$, while $\lambda(t)$ increases more
slowly than $t$, the Hubble radius increases faster than $\lambda(t)$, and a
length scale $\lambda(t)$ will start larger than the Hubble radius
($\lambda>\lambda_H$), cross the Hubble radius ($\lambda=\lambda_H$), and then
will remain inside the Hubble radius ($\lambda<\lambda_H$).

The story is different if we imagine that the Universe was once in an
inflationary phase. In inflation $H$ is roughly constant, so the Hubble length
is roughly time independent.  Thus, a given scale can start sub-Hubble radius,
$\lambda<\lambda_H$, then pass outside the Hubble radius during inflation, and
then re-enter the Hubble radius after inflation.  Thus, perturbations can be
imparted on a given length scale in the inflationary era as that scale leaves
the Hubble radius, and will be present as that scale re-enters the Hubble
radius after inflation in the radiation-dominated or matter-dominated eras.

Microphysics cannot affect the perturbation while it is outside the Hubble
radius, and the evolution of its amplitude is {\em kinematical}, unaffected
by dissipation, the equation of state, instabilities, and the like. However,
for super-Hubble radius sized perturbations one must take into account the
freedom in the choice of the background reference space-time, i.e., the gauge
ambiguities.  As usual when confronted with such a problem, it is convenient
to calculate a {\em gauge-invariant} quantity.  For inflation it is
convenient to study a quantity conventionally denoted $\zeta$ \re{BARDEEN}. In
the uniform Hubble constant gauge, at Hubble radius crossing $\zeta$ is
particularly simple, related to the background energy density and pressure
$\rho_0$ and $p_0$, and the perturbed energy density $\rho_1$:
\be
\label{eq:ZETA}
\zeta\equiv \delta \rho/(\rho_0+p_0),
\ee
where $\delta\rho=\rho_1-\rho_0$ is the density perturbation.

In the standard matter-dominated (MD) or radiation-dominated (RD) phase,
$\zeta$ at Hubble radius crossing (up to a factor of order unity) is equal to
$\delta\rho /\rho$.  Thus, the amplitude of a density perturbation when it
crosses back inside the Hubble radius after inflation,
$(\delta\rho /\rho)_{\rm HOR}$,\footnote{The notation ``HOR'' follows because
often in the literature the Hubble radius is referred to (incorrectly) as the
horizon.} is given by $\zeta$ at the time the fluctuation crossed outside the
Hubble radius during inflation.

As inferred from the adoption of $\zeta$, the convenient specification of the
amplitude of density perturbations on a particular scale is when that
particular scale just enters the Hubble radius, denoted as
$(\delta\rho /\rho )_{\rm HOR}$. Specifying the amplitude of the perturbation
at Hubble radius crossing evades the subtleties associated with the gauge
freedom, and has the simple Newtonian interpretation as the amplitude of the
perturbation in the  gravitational potential. Of course, when one specifies
the fluctuation spectrum at Hubble radius crossing, the amplitudes for
different lengths are specified at {\em different} times.

Now let us turn to the scalar field dynamics during inflation.  Consider a
minimally coupled, spatially homogeneous scalar field $\phi$, with Lagrangian
density
\be
{\cal L} = \partial^\mu\phi\partial_\mu\phi /2 -V(\phi )
= \dot\phi^2/2 - V(\phi ).
\ee
With the assumption that $\phi$ is spatially homogeneous, the stress-energy
tensor takes the form of a perfect fluid, with energy density and pressure
given by $\rho_\phi = \dot\phi^2/2+V(\phi )$, and $p_\phi = \dot\phi^2/2-V(\phi
)$. The classical equation of motion for $\phi$ is
\be
\label{eq19b}
\ddot\phi +3H\dot\phi + V^\prime (\phi ) = 0,
\ee
with the expansion rate in a flat FRW spacetime given by
\be
\label{eq19a}
H^2 = \frac{\kappa^2}{3} \, \left( \frac{1}{2} \dot{\phi}^2 +V(\phi) \right).
\ee
Here dot and prime denote differentiation with respect to cosmic time and
${\phi}$ respectively.  We assume that inflation has already provided us with a
flat universe by the time the largest observable scales cross the Hubble
radius.

By differentiating Eq.\ (\ref{eq19a}) with respect to $t$ and substituting in
Eq.\ (\ref{eq19b}), we arrive at the ``momentum'' equation
\be
2\dot{H} = -\kappa^2 \dot{\phi}^2.
\ee
All minimal slow-roll models are examples of sub-inflationary behavior, which
is defined by the condition $\dot{H} <0$. Super-inflation, where $\dot{H}>0$,
cannot occur here, though it is possible in more complex scenarios
[\ref{LPLB}, \ref{POWER}]. We may divide both sides of this equation by
${\dot{\phi}}$ if this quantity does not pass through zero.  This allows us to
eliminate the time-dependence in the Friedmann equation [Eq.\ (\ref{eq19a})]
and derive the first-order, non-linear differential equations
\bea
\label{eq1a}
(H')^2-\frac{3}{2}\kappa^2 H^2 & = & -\frac{1}{2} \kappa^4 V(\phi)\\
\label{eq1b}
\kappa^2 \dot{\phi} & = & -2H' .
\eea

A common framework for discussion of inflation is the slow-roll approximation,
though let us emphasize here that in much of our treatment of inflaton dynamics
we shall not need to resort to it. We can define two parameters, which we will
denote as slow-roll parameters, by\footnote{These definitions differ slightly
from, and indeed improve upon, those of Refs.\ (\ref{ST1984}), (\ref{LL})
which were made using the potential rather than the Hubble parameter. As
defined here they possess rather more elegant properties.}
\bea
\label{eq:ASD}
\epsilon & \equiv &  \frac{3\dot{\phi}^2}{2} \left( V +
	\frac{\dot{\phi}^2}{2} \right)^{-1}  = \frac{2}{\kappa^2} \,
	\left( \frac{H'}{H} \right)^2 \nonumber \\
\eta & \equiv & \frac{\ddot{\phi}}{H\dot{\phi}}= \frac{2}{\kappa^2} \,
	\frac{H''}{H} .
\eea
Slow-roll corresponds to $\{\epsilon, |\eta| \}  \ll 1$. These conditions
correspond respectively to the cases when the first term in
Eq.\ (\ref{eq1a}) and the first term in its $\phi$-derivative can be neglected.

With these definitions, the end of inflation is given {\em
exactly}\footnote{With the definition of $\epsilon$ in Refs.\ (\ref{ST1984}),
(\ref{LL}), this result is true only in the slow-roll approximation.} by
$\epsilon = 1$. A small value of $\eta$ guarantees
\be
\label{eq:SLO}
3H\dot{\phi} \simeq -V^{\prime}(\phi),
\ee
which is often called the slow-roll equation.\footnote{Note the difference
between slow-roll inflation and the slow-roll equation.  Slow-roll inflation
is a model where inflation occurs where the scalar field is slowly evolving to
its minimum, while the slow-roll equation implies that $ \ddot{\phi}$ can be
neglected.} Although the terminology ``slow-roll approximation'' is normally
used rather loosely, one could imagine carrying out a formalized perturbation
expansion in the slow-roll parameters, and we shall refer to such results
later.

Density perturbations arise as the result of quantum-mechanical fluctuations
of fields in de Sitter space.  First, let's consider scalar density
fluctuations.  To a good approximation we may treat the inflaton field $\phi$
as a massless, minimally coupled field. (Of course the inflaton does have a
mass, but inflation operates when the field is evolving through a ``flat''
region of the potential.) Just as fluctuations in the density field may be
expanded in a Fourier series as in Eq.\ (\ref{eq:TWOONE}), the fluctuations in
the inflaton field may be expanded in terms of its Fourier coefficients
$\delta\phi_{{\bf k}}$: $\delta\phi({\bf x})\propto\int\delta \phi_{{\bf k}}
\exp(-i{\bf k}\cdot{\bf x})d^3\!k$.  During inflation there is an event
horizon as in de Sitter space, and quantum-mechanical fluctuations in the
Fourier components of the inflaton field are given by \re{BUNCHDAVIES}
\be
\label{eq:LKJ}
k^3\left|\delta\phi_{{\bf k}}\right|^2/2\pi^2 = (H/2\pi)^2,
\ee
where $H/2\pi$ plays a role similar to the Hawking temperature of black holes.
Thus, when a given mode of the inflaton field leaves the Hubble radius during
inflation, it has impressed upon it quantum mechanical fluctuations.  In
analogy to Eq.\ (\ref{eq:DHROL}), what is called the fluctuations in the
inflaton field on scale $k$ is proportional to $k^{3/2}|\delta\phi_{{\bf k}}|$,
which by Eq.\ (\ref{eq:LKJ}) is proportional to $H/2\pi$.  Fluctuations in
$\phi$ lead to perturbations in the energy density:
\be
\delta\rho_\phi=\delta\phi(\partial V/\partial\phi).
\ee
Now considering the fluctuations as a particular mode leaves the Hubble radius
during inflation, we may construct the gauge invariant quantity $\zeta$ from
Eq.\ (\ref{eq:ZETA}) using the fact that during inflation
$\rho_0+p_0=\dot{\phi}^2$:
\be
\zeta = \delta\phi \left(\frac{\partial V}{\partial \phi}\right)
	\frac{1}{\dot{\phi}^2}.
\ee

Now using Eq.\ (\ref{eq1a}) and Eq.\ (\ref{eq1b}), the amplitude of the
density perturbation when it crosses the Hubble radius {\em after} inflation
is
\be
\label{eq:TYU}
\left(\frac{\delta \rho}{\rho}\right)_\lambda^{\rm HOR} \equiv
\frac{m}{\sqrt{2}} A_S(\phi) = \frac{m\kappa^2}{8\pi^{3/2}} \,
\frac{H^2(\phi)}{|H'(\phi)|} \propto \frac{V^{3/2}(\phi)}{m_{Pl}^3V'(\phi)} ,
\ee
where $H(\phi)$ and $H'(\phi)$ are to be evaluated when the scale $\lambda$
crossed the Hubble radius {\em during} inflation. The constant $m$ equals
$2/5$ or $4$ if the perturbation re-enters during the matter or radiation
dominated eras respectively.\footnote{The $4$ for radiation is appropriate to
the uniform Hubble constant gauge.  One occasionally sees a value $4/9$
instead which is appropriate to the synchronous gauge. The matter domination
factor is the same in either case. Note also that it is exact for matter
domination, but for radiation domination it is only strictly true for modes
much larger than the Hubble radius, and there will be corrections in the
extrapolation down to the size of the Hubble radius.}  Now we wish to know the
$\lambda$-dependence of $(\delta\rho/\rho)_\lambda$, while the right-hand side
of the equation is a function of $\phi$ when $\lambda$ crossed the Hubble
radius during inflation. We may find the value of the scalar field when the
scale $\lambda$ goes outside the Hubble radius in terms of the number of {\em
e}-foldings of growth in the scale factor between Hubble radius crossing and
the end of inflation.

It is quite a simple matter to calculate the number of $e$-foldings of growth
in the scale factor that occur as the scalar field rolls from a particular
value $\phi$ to the end of inflation $\phi_e$:
\be
\label{eq:NNN}
N(\phi)\equiv \int_{te}^{t} H(t') dt' =
- \frac{\kappa^2}{2} \int_{\phi }^{\phi_e} \frac{H(\phi')}{H'(\phi')} d\phi'.
\ee
The slow-roll conditions guarantee a large number of $e$-foldings. The total
amount of inflation is given by $N_{\rm TOT}\equiv N(\phi_i)$, where $\phi_i$
is the initial value of $\phi$ at the start of inflation (when $\ddot{a}$
first becomes positive). In general, the number of {\em e}-folds between when
a length scale $\lambda$ crossed the Hubble radius during inflation and the
end of inflation is given by \re{ST1984}
\be
\label{eq:NLAS}
 N (\lambda) = 45 +  \ln (\lambda /{\rm Mpc})
     +{2\over 3}\ln (M/10^{14}\,{\rm GeV})
         +{1\over 3}\ln (T_{\rm RH}/10^{10}\,{\rm GeV}),
\ee
where $M$ is the mass scale associated with the potential and $T_{\rm RH}$ is
the ``re-heat'' temperature. Relating $N(\lambda)$ and $N(\phi)$ from Eq.\
(\ref{eq:NNN}) results in an expression between $\phi$ and $\lambda$.
Hopefully this dry formalism will become clear in the example discussed below.

In addition to the scalar density perturbations caused by de Sitter
fluctuations in the inflaton field, there are gravitational mode
perturbations, $g_{\mu\nu}\rightarrow g_{\mu\nu}^{\rm FRW}+h_{\mu\nu}$, caused
by de Sitter fluctuations in the metric tensor [\ref{AW},\ref{STAR}].  Here,
$g_{\mu\nu}^{\rm FRW}$ is the Friedmann--Robertson--Walker metric and
$h_{\mu\nu}$ are the metric perturbations.  That de Sitter space fluctuations
should lead to fluctuations in the metric tensor is not surprising, since
after all, gravitons are the propagating modes associated with transverse,
traceless metric perturbations, and they too behave as minimally coupled
scalar fields.  The dimensionless tensor metric perturbations can be expressed
in terms of two graviton modes we will denote as $h$.  Performing a Fourier
decomposition of $h$, $h(\exx)\propto\int\delta
h_k\exp(-i\kay\cdot\exx)d^3\!k$, we can use the formalism for scalar field
perturbations simply by the identification $\delta\phi_{{\bf k}} \rightarrow
h_{{\bf k}}/\kappa\sqrt{2}$, with resulting quantum fluctuations [cf. Eq.\
(\ref{eq:LKJ})]
\be
k^3|h_{{\bf k}}|^2/2\pi^2 = 2\kappa^2(H/2\pi)^2.
\ee

While outside the Hubble radius, the amplitude of a given mode remains
constant, so the amplitude of the dimensionless strain on scale $\lambda$ when
it crosses the Hubble radius after inflation is given by
\be
\label{eq:FDK}
\left|k^{3/2}h_{{\bf k}}\right|_\lambda^{\rm HOR} \equiv A_G(\phi) =
   \frac{\kappa}{4\pi^{3/2}}\, H(\phi) \sim \frac{V^{1/2}(\phi)}{m_{Pl}^2},
\ee
where once again $H(\phi)$ is to be evaluated when the scale $\lambda$ crossed
the Hubble radius {\em during} inflation.

As usual, it is convenient to illustrate the general features of inflation in
the context of the simplest model, chaotic inflation \re{CHAOTIC}, which is to
inflationary cosmology what {\em drosophila} is to genetics.  In chaotic
inflation the inflaton potential is usually taken to have a simple polynomial
form such as $V(\phi)=\lambda\phi^4$, or $V(\phi) = \mu^2 \phi^2$. For a
concrete example, let us consider the simplest chaotic inflation model, with
potential $V(\phi) = \mu^2 \phi^2$ \re{BGKZ}. This model can be adequately
solved in the slow-roll approximation, yielding
\bea
\phi(t) & = & \phi_i - \frac{2}{\sqrt{3}} \frac{\mu}{\kappa}t \nonumber \\
a(t) & = & a_i \exp \left[ \frac{\kappa \mu}{{\sqrt{3}}}  \left(
\phi_i t - \frac{\mu}{{\sqrt{3}} \kappa} \, t^2 \right) \right] \nonumber \\
H & = & \frac{\kappa\mu}{\sqrt{3}}
	\left(\phi_i-\frac{2}{\sqrt{3}}\frac{\mu}{\kappa}t
\right) = \frac{\kappa\mu}{\sqrt{3}}\phi,
\eea
with inflation ending at $\kappa\phi_e = \sqrt{2}$ as determined by
$\epsilon=1$, where $\epsilon$ was defined in Eq.\ (\ref{eq:ASD}). The number
of $e$-foldings between a scalar field value $\phi$, and the end of inflation
is just
\be
\label{eq:NPO}
N(\phi) = - \frac{\kappa^2}{2} \int_{\phi}^{\sqrt{2}/\kappa}
\frac{H(\phi')}{H'(\phi')} d\phi' = \frac{\kappa^2\phi^2}{4} - \frac{1}{2}.
\ee
Equating Eq.\ (\ref{eq:NPO}) and Eq.\ (\ref{eq:NLAS}) relates $\phi$ and
$\lambda$ in this model for inflation:
\be
\kappa^2 \phi^2 / 4 = \left[ 45.5 + \ln(\lambda/{\rm Mpc})\right].
\ee
Using Eq.\ (\ref{eq:TYU}) and Eq.\ (\ref{eq:FDK}), $A_S$ and $A_G$ are found
to be
\bea
A_S(\lambda) & = & \left(\sqrt{2}\kappa\mu/\sqrt{12\pi^3}\right)
	\left[ 45.5 + \ln(\lambda/{\rm Mpc})\right] \nonumber \\
A_G(\lambda) & = & \left(\kappa\mu/\sqrt{12\pi^3}\right)
	\left[ 45.5 + \ln(\lambda/{\rm Mpc})\right]^{1/2}.
\eea

We can note three features that are common to a large number of (but not all)
inflationary models. First, $A_S$ and $A_G$ have different functional
dependences upon $\lambda$.  Second, $A_G$ and $A_S$ increase with $\lambda$.
Finally, $A_S>A_G$, for scales of interest, although not by an enormous
factor.

To conclude this exercise, it is worth reminding the reader how little of the
inflaton potential is available for reconstruction. The scales of cosmological
interest at the present epoch lie in the range $1h^{-1}$ Mpc for galaxies up
to the current horizon size of $6000h^{-1}$ Mpc, where as usual $h$ is
Hubble's constant in units of 100 km s$^{-1}$Mpc$^{-1}$.  Taking the present
horizon distance to have crossed the Hubble radius $60$ $e$-foldings from the
end of inflation, we see that we only sample the small region of the potential
$V(\phi)$ for $\phi \in [2.7 m_{Pl}, 3.0 m_{Pl}]$.   By any standards, the
density perturbations from inflation we can actually sample represent an
extremely small region of the potential.  However it should be realized that
although we have potential information about a small region of the potential,
any information about the GUT potential, no matter how meager, is precious!
Indeed, in the exploration of GUTs, cosmology may reveal the first ``piece of
the action.''

We have one further piece of information, which is that we know that the
remainder of inflation must occur in the remaining section of the potential,
with the scalar field coming to rest with $V(\phi) = 0$. Although this
represents a significant constraint on the potential on scales below those
that large-scale structure observations can sample, it does still leave an
uncountable infinity of possible forms in this region. (One other constraint
in this region comes from primordial black holes, whose abundance can in
principle be calculated from the spectrum. Should black hole formation be
copious, this constrains the spectrum at the mass scales corresponding to the
size of detectable black holes, which are most prominent at around
$10^{15}$g \re{CL1993}.)

To conclude this section we call the reader's attention to Figure 1, which
illustrates the procedure.  The figure illustrates a scalar field $\phi$
rolling down a potential $V(\phi)$.  At some point in the evolution the
slow-roll conditions break down and inflation ends.  We can count back from
this point the number of $e$-foldings from the end of inflation, and use this
information to find a  relationship between the value of $\phi$ in the
evolution, and the length scale $\lambda$ leaving the Hubble radius at that
point.  While $\phi$ evolves, quantum fluctuations imprint scalar and
gravity-wave perturbations upon each scale as it leaves the Hubble radius. The
scalar perturbations depend upon the potential and its derivative, while the
gravitational modes depend only upon the potential.  In principle, $A_G$ and
$A_S$ are probed by observations of large-scale structure and by measurements
of CMBR fluctuations.  The length scale, and corresponding angular scales, of
several important observations are indicated.

In the next section we will discuss the procedure for reversing the process
discussed in this section; knowing $A_G(\lambda)$ and $A_S(\lambda)$, how does
one determine $V(\phi)$?

\vspace{48pt}
\thesection{\centerline{\bf III. RECONSTRUCTION OF THE POTENTIAL}}
\setcounter{section}{3}
\setcounter{equation}{0}
\vspace{18pt}

A number of authors [\ref{LL},\ref{TENSORS},\ref{DHSST},\ref{LMM}] have
recently emphasized the possibility that tensor modes excited during
inflation, corresponding to gravitational waves, may play an important role in
generating microwave background anisotropies. We thus develop an extension of
the potential reconstruction methods of Hodges and Blumenthal \re{HB} to
include tensor as well as scalar modes. As discussed in the previous section,
the expressions for the amplitudes of the scalar and tensor modes may be
written as
\bea
\label{eq2}
A_S(\phi) & = & \frac{\sqrt{2}\kappa^2}{8\pi^{3/2}} \,
	\frac{H^2(\phi)}{|H'(\phi)|}\nonumber \\
A_G(\phi) & = & \frac{\kappa}{4\pi^{3/2}}\, H(\phi),
\eea
respectively.  Note that the definition of $A_S(\phi)$ in Eq.\ (\ref{eq2}) is
related to the power spectrum $P^{1/2}(k)$ defined in Hodges and Blumenthal
\re{HB} by
\be
P^{1/2}(k) = 3 \sqrt{2\pi} A_S(\phi).
\ee

Utilizing the slow-roll approximation, there are useful expressions for the
scale-dep\-end\-ence of the spectra, the {\em spectral indices}, to
first-order in departure from slow-roll. These are
\bea
1-n  \equiv &   d \ln [A_S^2 (\lambda)]/ d \ln \lambda & =
	4\epsilon_* - 2 \eta_* \nonumber\\
n_{{\scriptscriptstyle G}} \equiv &    d \ln [A_G^2 (\lambda)]/ d \ln
	\lambda & = 2 \epsilon_*
\eea
where ``$*$'' indicates evaluation at the time when the scale $\lambda$ passes
outside the Hubble radius during inflation. In keeping with convention we drop
the subscript $S$ on the scalar mode index. Whenever the slow-roll conditions
are closely obeyed, the spectrum is close to scale invariant. When this is not
true, there are in general corrections to the expressions for the fluctuations
at the next order in an expansion in slow-roll parameters.

The reader may have noticed that although we are keeping the equations of
motion general ({\it i.e.,} not subject to a slow-roll approximation), our
expression for the scalar modes in Eq.\ (\ref{eq2}) is an expression based on
the slow-roll approximation, $\{\epsilon,|\eta|\} \ll 1$.  Ideally, one would
like to completely abandon the slow-roll regime, because within it, the scalar
spectrum is always close to the scale-invariant case and the gravitational
wave amplitude is always small, as we have seen. In practice, it seems very
possible that should inflation have occurred, it may well have been pushing
the outside of the slow-roll approximation envelope, and indeed much of the
recent interest has been in the possibilities of both tilt and gravitational
waves. True reconstruction assumes nothing about $V(\phi)$ (flatness, etc.)
except that it inflates. Unfortunately, although we are able to keep the
dynamics completely general, general expressions are not available for the
perturbation spectra.

Recently, an improvement has become available in the form of general
expressions for the spectra to first-order in departure from slow-roll
\re{LS93}. These give rise to ``first-order corrected'' spectra, which can be
written
\bea
A_S^{{\rm corr}} & = & [1- \epsilon + (2 - \ln 2 - \gamma)(2\epsilon-\eta)]
	A_S^{{\rm uncorr}} \nonumber \\
A_G^{{\rm corr}} & = & [1+ (1 - \ln 2 - \gamma)\epsilon] A_G^{{\rm uncorr}},
\eea
where $\gamma \simeq 0.577$ is Euler's constant. If slow-roll is breaking then
these can represent a significant improvement on the uncorrected results, but
unfortunately the reconstruction loses its analytic tractability. The one
exception to this is the case of power-law inflation ---in that case the
effects of the corrections cancel exactly \re{LS} in the reconstruction
equation Eq.~(\ref{eq10}) we derive below.

Rather than resort immediately to numerical construction, we elect instead to
make the operational choice that we shall adopt the slow-roll expressions for
the spectra. A reconstruction can then be made subject to a consistency check
that the slow-roll conditions are indeed satisfied; if not, then our formalism
will have to be enhanced to incorporate these improvements.

It is clear that the ratio of amplitudes of the scalar and tensor modes is
given by
\be
\label{extra}
\frac{A_G}{A_S}=\frac{\sqrt{2}}{\kappa}\, \frac{|H'|}{H}= \sqrt{\epsilon},
\ee
and if $\epsilon \ll 1$, then $A_G/A_S \ll 1$. It is possible that the {\em
COBE}
satellite is in fact observing a sum of contributions from the tensor and
scalar fluctuations, as opposed to the pure scalar modes as originally
thought. If these are uncorrelated and obey Gaussian statistics, the quantity
of observational interest on large angular scales is the sum of the
squares\footnote{The relative weighting of $A_S^2$ and $A_G^2$ in this
equation is that appropriate to large angle anisotropies (greater than
$2^\circ$) in the slow-roll approximation. This is discussed in depth in
Section VI, and exact weighting formulae provided there.}
\be
S^2(\phi)=\frac{m^2}{2}A_S^2(\phi)+A_G^2(\phi).
\ee
Using $m=2/5$ and recalling that $\epsilon$ must be less than unity, we see
immediately that the tensor modes dominate $S^2(\phi)$ if $2/25<\epsilon<1$, or
equivalently if ${\kappa}/5<|H'|/H<{\kappa}/{\sqrt{2}}$.  The largest relative
tensor contribution to $S^2(\phi)$ obtains for $\epsilon=1$: $2A_G^2/m^2 A_S^2
= 25/2$ for $m=2/5$.\footnote{If one is performing a theoretical
reconstruction of the potential by specifying either $A_G$ or $A_S$, it is
essential to ensure this condition is always satisfied for consistency.
Indeed, observations violating $A_G/A_S<1$ would immediately rule out the
models we are considering.}  Although it is not mandatory (one can break
slow-roll only in the $\eta$ parameter as in natural inflation), the
gravitational wave contribution is typically significant whenever there is a
deviation from the slow-roll regime.

On the other hand, the gravitational waves behave as relativistic matter when
they re-enter the Hubble radius and do not interact with the other matter
components. Consequently their energy density redshifts as $a^{-4}$, which
implies that only scalar modes affect the CMBR anisotropy on angular scales
$\theta \ll 2^\circ$. However the anisotropy on these scales is also affected
by the form of dark matter present. [For a recent discussion of some of these
issues, see Ref.\ (\ref{T1993}).]

To proceed, we shall assume that the functional forms of $A_S(\lambda)$ and
$A_G(\lambda)$ are known explicitly and defer until Section VI a discussion on
the many difficulties associated with determining these quantities from
observation. Our initial aim is to develop a framework which allows the
inflaton potential to be determined. We consider general inflationary
behavior for the field equations (\ref{eq1a}) and (\ref{eq1b}) and it proves
convenient to parameterize the full set of solutions in terms of the function
$H({\lambda}(\phi))$, where ${\lambda}$ is the scale. Eqs.\ (\ref{eq2}) now
become
\bea
\label{eq6a}
A_S(\lambda) & = & \frac{\sqrt{2}\kappa^2}{8\pi^{3/2}} \, H^2(\lambda) \left|
\frac{d{\lambda}}{dH}\, \frac{d\phi}{d\lambda}
\right| \nonumber \\
A_G(\lambda) & = & \frac{\kappa}{4\pi^{3/2}} \, H(\lambda).
\eea

Each length scale $\lambda$ is associated with a unique value of $\phi$ when
that scale crossed the Hubble radius during inflation.  We will indicate that
relationship by writing $\lambda(\phi)$.  Now when a present length scale
$\lambda$ crossed the Horizon radius during inflation with scalar field value
$\phi$, its physical size was $H^{-1}(\phi)$.  The physical size grew between
horizon crossing and today, and is now simply
$\lambda(\phi)=H^{-1}(\phi)a_0/a(\phi)$, where $a_0$ is the present value of
the scale factor and $a(\phi)$ was the value of the scale factor when the scale
crossed the Hubble radius during inflation.   Now we can make use of Eq.\
(\ref{eq:NNN}) to relate $a(\phi)$ to the value of the scale factor at the end
of inflation, $a_e$: $a(\phi)=a_e\exp[-N(\phi)]$.  This allows us to express
$\lambda(\phi)$ as
\be
\label{eq8}
\lambda(\phi)=\frac{\exp[N(\phi)]}{H(\phi)} \, \frac{a_0}{a_e},
\ee
where $N(\phi)$ is given by Eq.\ (\ref{eq:NNN}). Differentiating Eq.\
(\ref{eq8}) with respect to ${\phi}$ yields
\be
\label{eq9}
\frac{d\lambda(\phi)}{d\phi}= \pm \frac{\kappa}{\sqrt{2}} \left(
\frac{A_S}{A_G} - \frac{A_G}{A_S} \right) \lambda ,
\ee
and taking the ratio of Eqs.\ (\ref{eq6a}) implies\footnote{If one were to use
the `first-order corrected' expressions for the spectra discussed earlier, the
right hand side of Eq.~(\ref{eq10}) would be multiplied by $(1-1.27 \epsilon +
1.27 \eta)$.}
\be
\label{eq10}
\frac{\kappa}{\sqrt{2}}\, \frac{A_G}{A_S} = \left| \frac{d\lambda}{d\phi} \,
	\frac{d\ln A_G}{d\lambda} \right| .
\ee
Note that expression (2.10) in Hodges and Blumenthal \re{HB} consists of only
our first term in Eq.\ (\ref{eq9}), indicating their assumption of slow-roll
behavior. Substituting Eq.\ (\ref{eq9}) into Eq.\ (\ref{eq10}) gives
\be
\label{eq:NEWSTUF}
\frac{\lambda}{A_G(\lambda)}\, \frac{dA_G(\lambda)}{d\lambda} =
\frac{A_G^2(\lambda)}{A_S^2(\lambda)-A_G^2(\lambda)}.
\ee
Note that the left hand side is just equal to $n_G/2$. This equation is
similar to Eq.~(9) in Davis {\it et al.} \re{DHSST}, provided one interprets
their $n$ as being the tensor index and not the scalar one. It clearly shows
that there exists a correspondence between the scalar and tensor modes and is
valid for an arbitrary interaction potential. In principle, if the
scale dependence of either the scalar {\em or} tensor modes is known, the
other can be determined from Eq.\ (\ref{eq:NEWSTUF}). If only $A_G(\lambda)$
is known, then $A_S(\lambda)$ follows immediately by differentiation. However,
if only $A_S(\lambda)$ is known, a first-order differential equation must be
solved to find the form of $A_G(\lambda)$. Thus, knowledge of only the scalar
spectrum leaves an undetermined constant in the tensor spectrum.

Once the form of the tensor spectrum is known, the potential, as parametrized
by $\lambda$, may be derived by substituting Eqs.\ (\ref{eq6a}) into Eq.\
(\ref{eq10}). We find
\be
\label{eq13}
V[\phi(\lambda)] = \frac{16\pi^3A_G^2(\lambda)}{\kappa^4}\left[ 3 -
	\frac{A_G^2(\lambda)}{A_S^2(\lambda)} \right] .
\ee
Finally, integration of Eq.\ (\ref{eq9}) yields the function
$\phi=\phi(\lambda)$ given by
\be
\label{eq14}
\phi(\lambda) =  \pm \frac{\sqrt{2}}{\kappa } \int^\lambda
	\frac{d\lambda'}{\lambda'}
	\frac{A_S(\lambda')A_G(\lambda')}{A_S^2(\lambda')-A_G^2(\lambda')}.
\ee
We have absorbed the integration constant by taking advantage of the freedom
to shift $\phi$ by a constant.  The functional form of $V(\phi)$ follows by
inverting Eq.\ (\ref{eq14}) and substituting the result into Eq.\
(\ref{eq13}).   It will also prove convenient at times to express $\phi$ in
terms of $A_G$.  If the functional form of $A_S$ as a function of $A_G$ is
known, $A_S[A_G]$, then using Eq.\ (\ref{eq:NEWSTUF}) in Eq.\ (\ref{eq14})
gives
\be
\label{eq:STV}
\phi=\pm \frac{\sqrt{2}}{\kappa} \int^{A_G} dA_G'\, \frac{A_S[A_G']}{A_G'^2 }.
\ee

The reconstruction equations are Eqs.\ (\ref{eq:NEWSTUF}), (\ref{eq13}), and
(\ref{eq14}). It is worth emphasizing again that for any choice of
$A_G(\lambda)$, there is a unique associated $A_S(\lambda)$ and $V(\phi)$ (at
least in the slow-roll approximation), but that the converse is not true. As
shown by Hodges and Blumenthal \re{HB},  the scalar spectrum leaves an
undetermined constant in the tensor spectrum, and as the equation relating $V$
and $A_G$ is non-linear, different choices of this constant might lead to
functionally different forms of the potential \re{LT1993}. In order to
reconstruct the potential from scalar modes, one needs an additional piece
of information. Technically what is needed is knowledge of the functional
dependence of $A_S$ upon $A_G$, $A_S[A_G]$.  This can be fixed either by
knowledge of the amplitude of the tensor spectrum at a single scale, which
would fix $A_G$ uniquely, or knowledge that $A_S$ is independent of $A_G$. As
$A_G$ cannot be independent of $\lambda$, the latter possibility arises only
if $A_S(\lambda)$ is constant.

It is also worth emphasising consistency, which can provide an important
check. If our inflationary assumptions are correct, then the two spectra are
intimately related as illustrated above. However, observations are typically
subject to both systematic and statistical errors, and within these one might
find that measured spectra are not exactly consistent. Were one to be
confronted with such data, one would like some prescription by which to decide
how to best reconcile the data, to generate some kind of ``maximum likelihood''
reconstruction. Such a procedure  would presumably also allow one to
demonstrate that the measured spectra were not compatible with each other
within the inflationary paradigm, if indeed inflation were not the correct
source of the fluctuations. In practice, the situation is skewed by the scalar
fluctuations being considerably easier to observe than their tensor
counterparts, and it seems prudent to await the arrival of considerably better
data before properly contemplating how one would deal with the possibility of
marginally incompatible observations.

The reconstruction procedure simplifies if $A_G(\lambda)\ll A_S(\lambda)$
(i.e., $\epsilon\ll 1$):
\bea
\label{eq:AGLAS}
\frac{\lambda}{A_G(\lambda)} \frac{dA_G(\lambda)}{d\lambda} & = &
\frac{A_G^2(\lambda)}{A_S^2(\lambda)} \nonumber \\
\phi(\lambda) & = &
	\pm \frac{\sqrt{2}}{\kappa}\int^\lambda \frac{d\lambda'}{\lambda'}\,
\frac{A_G(\lambda')}{A_S(\lambda')} \nonumber \\
V[\phi(\lambda)] & = & \frac{48\pi^3}{\kappa^4} A_G^2(\lambda).
\eea

We conclude this section by summarizing the conditions necessary for the
perturbation amplitudes to increase or decrease with increasing wavelength.
Such information alone can place strong limits on the functional form of the
potential. The scales that first cross the  Hubble radius are the last
to re-enter during the radiation or matter dominated eras (see Fig.\ 1).
Consequently, the amplitudes of the modes increase (decrease) with wavelength
if they decrease (increase) with time during inflation. Immediately we
conclude that
\be
\label{eq15}
\frac{dA_G}{d\lambda} > 0
\ee
for all sub-inflationary ($\dot{H}<0$) models. One requires an era of
super-inflation ($\dot{H}>0$) if this inequality is to be reversed.
Super-inflation is only possible with a minimally coupled self-interacting
scalar field if the spatial hypersurfaces of the manifold have
positive-definite curvature \re{LPLB}. An observation indicating
$dA_G/d\lambda < 0$ would therefore require some of the main assumptions made
in the inflationary analysis to be significantly altered. Within the context
of the FRW Universe, for example, one would need to extend the gravitational
sector of the theory beyond general relativity, or assume that the value of
the density parameter was significantly larger than unity at first Hubble
radius crossing. Indeed, Eq.\ (\ref{eq15}) implies that any effects of the
gravitational waves on the CMBR anisotropy will always be enhanced on larger
angular scales in the models considered here.

On the other hand, it is possible for the scalar spectrum to decrease with
wavelength. By writing $dA_S/d\lambda=(dA_S/d\phi)(d\phi/d\lambda)$ and
employing Eqs.\ (\ref{eq6a}) and (\ref{eq9}), one finds that a necessary and
sufficient condition for scalar modes to be decreasing in amplitude with
increasing wavelength is
\be
\label{eq16}
\left( \frac{H'}{H} \right)^2 < \frac{1}{2} \, \frac{H''}{H} .
\ee
In terms of the slow-roll parameters, this can be written as $2 \epsilon <
\eta$. As $\epsilon$ is positive by definition, this condition is not easy to
satisfy, particularly in the late stages of inflation where $\epsilon$ must
increase towards unity. A necessary, but not sufficient, condition for Eq.\
(\ref{eq16}) to hold is that the potential be {\em convex}, $V'' > 0$.
Therefore, if the field is located near a local maximum of the potential, as in
natural inflation \re{NAT} for example, the amplitude will always increase with
$\lambda$.

In conclusion, it is clear that any scale dependence for the spectrum of
gravitational waves is possible in principle, subject to condition
(\ref{eq15}). Secondly, the most useful parameter {\em mathematically} in the
reconstruction process is $A_G(\lambda)$, because once this is known the
potential can be derived in a rather straightforward manner.

\vspace{48pt}
\thesection{\centerline{\bf IV. RECONSTRUCTING THE FULL POTENTIAL}}
\setcounter{section}{4}
\setcounter{equation}{0}
\vspace{18pt}

Before proceeding  to analyze the possibilities for obtaining the spectra
observationally, we shall first illustrate some examples in reconstruction in
order to demonstrate the techniques. We shall examine four cases of increasing
complexity. These four cases will reconstruct to familiar potentials.

\vspace{18pt}

\centerline{\bf A. Polynomial potentials}

\vspace{18pt}

Let us first reconstruct the $\mu^2\phi^2$ chaotic potential model worked out
in Section II.  We will then generalize the result for construction of
polynomial potentials.

Recall that using the slow-roll approximation for the potential
$V(\phi)=\mu^2\phi^2$ we found perturbation spectra $A_G(\lambda) =
\alpha[\beta + \ln(\lambda/\lambda_0)]^{1/2}$ and $A_S(\lambda) =
\sqrt{2}A_G^2(\lambda)/\alpha$, with $\alpha^2=\kappa^2\mu^2/12\pi^3$,
$\beta=45.5$, and $\lambda_0=1$ Mpc.  We must keep in mind that these
solutions were obtained in the slow-roll approximation. Since the slow-roll
approximation implies that $A_G\ll A_S$, we must reconstruct using Eqs.\
(\ref{eq:AGLAS}).

First, let us reconstruct assuming that observations provide two pieces of
information: $A_G(\lambda)$ is of the form $A_G(\lambda) = \alpha[\beta
+\ln(\lambda/\lambda_0)]^{1/2}$, and  $A_G(\lambda)\ll A_S(\lambda)$.  Then
the differential equation for $d A_G(\lambda)/d\lambda$ in Eq.\
(\ref{eq:AGLAS}) can be used to yield a unique scalar spectrum,
$A_S(\lambda)=\sqrt{2}A_G^2(\lambda)/\alpha$, as anticipated from the
calculation in Section II.  (Of course $A_S(\lambda)$ could be found without
the assumption that $A_G(\lambda)\ll A_S(\lambda)$, but it would be
different.)

Now that we know both $A_G(\lambda)$ and $A_S(\lambda)$, we can find
$\phi(\lambda)$ from the second equation in Eq.\ (\ref{eq:AGLAS}):
\be
\label{eq:ALB}
\beta+\ln(\lambda/\lambda_0) = \kappa^2\phi^2/4.
\ee

Finally, we can use the last equation in the slow-roll reconstruction
procedure to give
\be
V(\phi)= \frac{48\pi^3}{\kappa^4} A_G^2(\lambda) = \frac{48\pi^3}{\kappa^4}
\alpha^2[\beta+\ln(\lambda/\lambda_0)] =
\frac{12\pi^3}{\kappa^2}\alpha^2\phi^2.
\ee
Exactly as expected, the potential is of the form $V(\phi)=\mu^2\phi^2$, with
$\mu^2=12\pi^3\alpha^2/\kappa^2$.  Thus, we have successfully reconstructed
the potential.

We began with the assumption that $A_G(\lambda)$ is known. If we had started
with the assumption that the scalar spectrum is known and of the form $
A_S(\lambda) = \sqrt{2}\alpha[\beta+\ln(\lambda/\lambda_0)],$ the differential
equation for $A_G(\lambda)$ would give
\be
A_G^{-2}(\lambda) = \alpha^{-2}[\beta + \ln(\lambda/\lambda_0)]^{-1} + C,
\ee
where $C$ is arbitrary. Fixing $A_G(\lambda_0) = \alpha\beta^{1/2}$ fixes
$C=0$, and reconstruction would proceed exactly as before.  Other choices of
$C$ would lead to different potentials, with different predictions for $A_G$.

Now let's consider a slightly more general tensor mode spectrum:
$A_G=\alpha[\beta + \ln(\lambda/\lambda_0)]^\gamma$ with $\gamma=$ constant,
again with $A_G(\lambda)\ll A_S(\lambda)$.  The differential equation for
$dA_G(\lambda)/d\lambda$ gives
\be
A_S(\lambda)=\left[\alpha/\sqrt{\gamma}\right]\left[\beta+
\ln(\lambda/\lambda_0)\right]^{(2\gamma+1)/2}.
\ee

The solution for $\phi(\lambda)$ is the same as Eq.\ (\ref{eq:ALB}) with
$\kappa \rightarrow \kappa/\sqrt{2\gamma}$.  Using this in the reconstruction
of the potential gives
\be
V(\phi)= \frac{48\pi^3}{\kappa^4} A_G^2(\lambda) =
\frac{48\pi^3}{\kappa^{4(1-\gamma)}} \, \frac{\alpha^2}{(8\gamma)^{2\gamma}}
\phi^{4\gamma}.
\ee
An oft studied case is $\gamma=1$, which reconstructs to
$V(\phi)=\lambda\phi^4$ with scalar and tensor perturbations
\bea
A_G(\lambda) & = & \alpha\left[\beta+\ln(\lambda/\lambda_0)\right] \nonumber \\
A_S(\lambda) & = & \alpha\left[\beta+
	\ln(\lambda/\lambda_0)\right]^{3/2}.
\eea

\vspace{18pt}

\centerline{\bf B. Harrison--Zel'dovich potentials}

\vspace{18pt}

Let us now look at potentials which give rise to the Harrison--Zel'dovich
spectrum, $A_S(\lambda) =a_S={\rm constant}$. Such spectra are actually rather
unlikely; most inflationary models exhibit a decrease in amplitude with
decreasing scale which is significant now given the accuracy of observations.

We start reconstruction by considering the differential equation relating
$A_G$ and $A_S$ [Eq.\ (\ref{eq:NEWSTUF})]:
\be
\frac{\lambda}{A_G(\lambda)}\, \frac{dA_G(\lambda)}{d\lambda} =
\frac{A_G^2(\lambda)}{a_S^2-A_G^2(\lambda)},
\ee
which has solution
\be
\label{eq:FGH}
\ln(\lambda/\lambda_0) = -\frac{a_S^2}{2}\left(\frac{1}{A_G^2(\lambda)}
	-\frac{1}{A_{G0}^2}\right)-\ln(A_G(\lambda)/A_{G0}).
\ee
In general there is no closed-form expression for $A_G(\lambda)$.

We can reconstruct the potential in two steps.  Since $A_S$ is a constant, we
can find $A_G$ in terms of $\phi$ by Eq.\ (\ref{eq:STV}):
\be
A_G^2(\phi)=2 a_S^2/\kappa^2\phi^2.
\ee
Now we can substitute this into the equation for $V$ in Eq.\ (\ref{eq13}) to
give
\bea
V(\phi) & = & \frac{16\pi^3}{\kappa^4} a_S^2\left[3
\frac{1}{(\phi/\bar{\phi})^2} - \frac{1}{(\phi/\bar{\phi})^4}\right]
\eea
where $\bar{\phi}=\sqrt{2}/\kappa$.

It should be emphasized that this is the {\em only} inflaton potential which
leads to an {\em exactly} scale-invariant spectrum of scalar density
fluctuations.  It arises as a special case of  ``intermediate'' inflation
\re{INTINF}, where the scale factor expands as $a \propto \exp (t^f)$ with $0
< f < 1$; the above potential corresponds to choosing $f = 2/3$. In contrast,
the spectrum of gravitational waves is not scale invariant.  It is generally
true that inflation cannot lead to scalar and tensor perturbation spectra that
are {\em both} constant in $\lambda$. It is interesting to note that
potentials of this form arise when supersymmetry is spontaneously broken
\re{WIT}.

We can reinterpret these results in terms of the slow-roll parameters. It is
clear that to obtain a flat spectrum we require $2 \epsilon = \eta$, but
$\epsilon$ and $\eta$ are not determined separately. There are some
interesting limiting cases. If we allow $a_S$ to tend to infinity, this
corresponds to $\epsilon$ tending to zero. In this limit the potential becomes
flat, with its constant value being that which gives the desired gravitational
wave amplitude. As $a_S$ is reduced from infinity, $\epsilon$ increases away
from zero preserving $2 \epsilon = \eta$. Once $\epsilon$ becomes big enough,
there will be slow-roll corrections which destroy the flatness of the
spectrum. It is interesting to note that although slow-roll automatically
guarantees a spectrum which is close to flat, it is perfectly possible for a
spectrum close to flatness to arise when the slow-roll conditions are not well
obeyed.

These potentials, which exhibit little tilt but which can have substantial
gravitational waves, are also of interest in that they complete a square of
possible behaviors in different inflationary models, as shown in Table 1.
Indeed, such a model performs well on most large-scale structure data with the
exception of intermediate-scale galaxy clustering data.

\begin{table}
\begin{center}
\begin{tabular}{c||c|c}
  Scalar     & Small gravitational        & Large gravitational     \\
  Spectrum   & wave contribution          & wave contribution       \\
\hline
\hline
              &                           &                         \\
Nearly flat   & Polynomial                & Harrison-Zel'dovich     \\
spectrum      & potentials                & potential               \\
              &                           &                         \\
Tilted        & Hyperbolic                & Exponential             \\
spectrum      & potentials                & potentials              \\
\end{tabular}
\end{center}
\footnotesize {\hspace*{0.2in} Table 1: Possible behaviors for spectra in
several inflationary models.}
\end{table}

\vspace{18pt}

\centerline{\bf C. Exponential potentials}

\vspace{18pt}

Generalizing away from the flat scalar spectrum, the simplest (and possibly
most
likely) case is where the amplitudes have a simple power-law dependence,
\be
\label{as1}
A_S(\lambda) = a_S (\lambda/\lambda_0)^\nu, \qquad   \nu \ne 0 ,
\ee
where $a_S$ is a constant. The recent measurements from {\em COBE} \re{DMR}
alone provide the constraint $-0.3 < \nu < 0.2$ at the $1$-sigma level.
Incorporating specific choices of dark matter and including clustering data
allows one to do better; for instance in a cold dark matter model (CDM) it has
been shown \re{LL2} that $\nu < 0.15$ at $95\%$ confidence in models with no
gravitational waves, and $\nu < 0.08$ (again $95\%$ confidence) in power-law
inflation which does have significant gravitational wave production.

$A_G(\lambda)$ satisfies the differential equation Eq.\ (\ref{eq:NEWSTUF})
\be
\label{gen1}
\frac{\lambda}{A_G(\lambda)}\, \frac{dA_G(\lambda)}{d\lambda} =
\frac{A_G^2(\lambda)}{a_S^2(\lambda/\lambda_0)^{2\nu}-A_G^2(\lambda)}.
\ee
Obtaining the general form for $A_G(\lambda)$ is difficult. However there are
some specific solutions which are of interest in that they relate to known
examples of inflationary potentials. One obvious solution to Eq.\ (\ref{gen1})
is $A_G(\lambda) = a_g (\lambda/\lambda_0)^{\nu}$, with
\be
\label{ag2}
a_g^{2} = \left( \frac{\nu}{1+\nu} \right) a_S^{2}, \qquad \nu \ne 0 .
\ee
Note that in this simple case, ${A_G}/{A_S} = {a_g}/{a_S}$, a constant
independent of scale, but that as $\nu \rightarrow 0$ the magnitude of the
tensor contribution reduces significantly.  We can simply integrate Eq.\
(\ref{eq14}) to obtain
\be
\phi(\lambda)= \pm \sqrt{ \frac{2}{\kappa^2} }
\ln \left( \frac{\lambda}{\lambda_0} \right)\sqrt{\nu^2+\nu} .
\ee
Substituting this expression into Eq.\ (\ref{eq13}) gives the final result
\be
\label{pot2}
V(\phi)=V_0\exp(\pm \phi/\bar{\phi}) ,
\ee
with
\be
\label{eqphi}
V_0 = \frac{16 \pi^{3}a_g^{2}}{\kappa^4}\,\frac{2\nu+3}{\nu+1}; \qquad
\bar{\phi}^{-1}= \kappa\sqrt{ \frac{2\nu}{\nu+1}} .
\ee

Thus we see that a power-law behavior for the amplitude of the scalar and
tensor modes is obtained from an exponential potential, and is therefore
consistent with power-law models of inflation \re{POWER}.

It is interesting to note that this result for $\bar{\phi}$ coincides with the
exact result for power-law inflation, whereas if slow-roll were strictly
applied one would get $\bar{\phi}^{-1} = \kappa\sqrt{2\nu}$, being the above to
lowest order in $\nu$. Thus our hybrid of general equations of motion but
slow-roll spectrum definitions certainly offers improved results over the
usual slow-roll method in this case.

Note that as $\nu \rightarrow + \infty$ the relative slope of the potential,
as determined by ${\alpha}$,  becomes independent of $\nu$. The limit $\alpha
(\nu = \infty)= {\sqrt{2}}\kappa$ corresponds to the Milne Universe $a(t)
\propto t$ and represents the limiting solution for inflation to occur. As
$\nu$ is increased in Eq.\ (\ref{pot2}) the only real effect is to increase
the height of the potential through the $V_0$ term.

Rather than the equal power behavior, consider the more general example
\be
\label{as2}
A_S(\lambda) = a_S (\lambda/\lambda_0)^\nu~;\qquad A_G(\lambda) =
	a_G (\lambda/\lambda_0)^\sigma,
\ee
where $a_S$ and $a_G$ are constants.   It is trivial to show that these
spectra are solutions to Eq.\ (\ref{eq:NEWSTUF}) or the differential equation
in Eq.\ (\ref{eq:AGLAS}) {\em only} if $\sigma=\nu$.  Thus, observation of
spectra that are exact power-laws with different powers would rule out the
class of inflationary models we consider as the source of the perturbations.

\vspace{18pt}

\centerline{\bf D. Hyperbolic potentials}

\vspace{18pt}

Let us return to the differential equation for $A_G(\lambda)$ in Eq.\
(\ref{gen1}), but in the $A_G(\lambda)\ll A_S(\lambda)$ limit.  The equation
becomes
\be
\frac{dA_G}{d\lambda} = \frac{A_G^3}{\lambda}\, \frac{1}{a_S^2}
	(\lambda/\lambda_0)^{-2\nu}.
\ee
This equation has general solution in terms of an undetermined constant
$\beta$:
\be
\label{eq:GENY}
A_G^2(\lambda) =a_S^2\nu \frac{(\lambda/\lambda_0)^{2\nu}}
	{1+\beta(\lambda/\lambda_0)^{2\nu}}.
\ee
We will see that different functional forms for the potential reconstruct
depending upon the sign of $\beta$.  Of course $\beta$ can be determined by
measurement of $A_G$ on any one scale.

As $\beta\rightarrow 0$ we recover the power-law spectra for $A_G(\lambda)$
and $A_S(\lambda)$ with equal power-law slopes.  This case was just considered
above. For small scales, $\left|\beta\right|(\lambda/\lambda_0)^{2\nu}\ll 1$,
we also recover the above case of equal power-law slopes for either choice of
the sign of $\beta$.  For $\beta>0$ we can take the limit of large scales,
$\beta(\lambda/\lambda_0)^{2\nu} \gg 1$, in which case $A_G(\lambda)$
asymptotically approaches a constant.

Recall that in the $A_G\ll A_S$ reconstruction procedure, Eq.\
(\ref{eq:AGLAS}) gives
\be
V[\phi(\lambda)] = \frac{48\pi^3}{\kappa^4}A_G^2(\lambda).
\ee
Now we must find $\phi(\lambda)$ and invert.

The integral expression for $\phi(\lambda)$ from Eq.\ (\ref{eq:AGLAS})
is
\be
\phi/\bar{\phi} = 2 \nu \int^\lambda \frac{d\lambda'}{\lambda'}
	\frac{1}{\left[1\pm|\beta|(\lambda'/\lambda_0)^{2\nu}\right]^{1/2}},
\ee
with ``$+$'' for positive $\beta$ and ``$-$'' for negative $\beta$.  The
constant $\bar{\phi}$ is the same as that of the previous subsection (in the
slow-roll approximation), $\bar{\phi}^{-1}= \kappa \sqrt{2 \nu}$. The
solutions to the integral are
\be
\phi/\bar{\phi} = \left\{ \begin{array}{ll}
	-{\rm 2 Arccsch}\left[\sqrt{\beta}\, (\lambda/\lambda_0)^{\nu}\right] &
	\beta>0 \\
	& \\
   	-{\rm 2 Arcsech}\left[\sqrt{|\beta|}\, (\lambda/\lambda_0)^{\nu}\right]
	& \beta<0 .
\end{array} \right.
\ee
These expressions are easily inverted to give $\lambda(\phi)$, and the
potential  reconstructs to
\be
\label{eq:HYPER}
V(\phi) = V_0 \left\{ \begin{array}{ll}
   \cosh^{-2}(\phi/2\bar{\phi}) & \beta>0 \\
                               & \\
   \sinh^{-2}(\phi/2\bar{\phi}) & \beta<0,
\end{array} \right.
\ee
with $V_0=48\pi^3a_S^2\nu/\kappa^4|\beta|$.

Note that for $\phi/\bar{\phi} \gg 1$, $V(\phi) \propto \exp(-\phi/
\bar{\phi})$ for both choices of the sign of $\beta$.   Large values of $\phi$
cross the Hubble radius late in inflation and correspond to small $\lambda$.
Notice from Eq.\ (\ref{eq:GENY}) that $A_G \rightarrow \lambda^\nu$ as
$\lambda \rightarrow 0$.  We have already reconstructed the potential that
results from this $A_G(\lambda)$ as $V(\phi /\bar{\phi})\propto\exp(-\phi /
\bar{\phi})$, which agrees with the definition of $\bar{\phi}$ given in Eq.
(\ref{eqphi}) when $\nu \ll 1$.  (The assumption that $A_G \ll A_S$ is
equivalent to this condition).

We can also expand Eq.\ (\ref{eq:HYPER}) for small $\phi$:
\be
V(\phi) \longrightarrow V_0 \left\{ \begin{array}{ll}
   1-  (\phi/\bar{\phi})^2 /4 + \cdots & \beta>0 \\
                                    & \\
   4 (\phi/\bar{\phi})^{-2} + \cdots          & \beta<0.
\end{array} \right.
\ee
The positive $\beta$ case is also an approximation to a potential of the form
$V(\phi)\propto 1+ \cos(\phi/\bar{\phi})$ as studied in a type of model called
natural inflation \re{NAT}.

The purpose of the above reconstruction exercises is to demonstrate how the
reconstruction process proceeds.  We have reconstructed several popular
inflationary potentials from knowledge of either the scalar or tensor
perturbation spectrum.  Before turning to the prospectus for actually
determining $A_S$ and $A_G$ from observational data, in the next section we
discuss a ``perturbative'' approach in reconstruction of the potential.

\vspace{48pt}
\thesection{\centerline{\bf V. RECONSTRUCTING A PIECE OF THE POTENTIAL}}
\setcounter{section}{5}
\setcounter{equation}{0}
\vspace{18pt}

The reconstruction program described in the previous section is quite
ambitious, as it depends upon knowledge of the functional forms of
$A_G(\lambda)$ and/or $A_S(\lambda)$ over a range of $\lambda$.  In this
section we will outline a less ambitious, but more realistic program.  We will
assume that we have information only about the scalar and tensor spectra (and
their first and second derivatives) at a single scale $\lambda_0$, and see
what we can learn about the potential.\footnote{In practice, observing the
derivatives at a single point may be just as difficult as measuring the shape
over a range of scales, though one might hope for adequate information to be
obtained from a significantly smaller range of scales (and with more freedom
to coarse-grain), perhaps even those accessible from a single experiment.}
This ``perturbative'' approach to reconstruction may be useful in the very
near future \re{T1993}.

\def\lz{{\lambda_0}}
\def\pz{{\phi_0}}
\def\vz{{V(\phi_0)}}
\def\vzp{{V'(\phi_0)}}

If we  know $A_G(\lz)$ and $A_S(\lz)$  at some length scale $\lz$ (which left
the Hubble radius during inflation when the value of the scalar field was
$\pz$),  we can use Eqs.\ (\ref{eq9}) and (\ref{eq14}) to find that
\bea
\label{eq:VB}
\frac{1}{\lambda} \left. \frac{d\lambda}{d\phi} \right|_{\lambda=\lz} & = &
\pm \frac{\kappa}{\sqrt{2}}\frac{A_S^2(\lz) - A_G^2(\lz)}{A_S(\lz)A_G(\lz)},
\nonumber \\
\left. \frac{dA_G(\lambda)}{d\lambda}\right|_{\lambda=\lz} & = &
\frac{1}{\lz}\, \frac{A_G^3(\lz)}{A_S^2(\lz)-A_G^2(\lz)}.
\eea
$\vz$ immediately follows from $A_G(\lz)$ and $A_S(\lz)$:
\be
\label{eq:VZ}
\vz = \frac{16\pi^3}{\kappa^4}\left[
3A_G^2(\lambda_0)-\frac{A_G^4(\lambda_0)}{A_S^2(\lambda_0)} \right].
\ee
With the approximation that $A_G(\lz) \ll A_S(\lz)$, the expression simplifies
to
\be
\vz = \frac{48\pi^3}{\kappa^4} A_G^2(\lambda_0) \left[ 1 + {\cal O} \left(
\frac{A_G^2(\lz)}{A_S^2(\lz)} \right) \right] \, .
\ee

Further reconstruction of the potential requires more than simply knowledge of
the amplitudes of the scalar and tensor perturbations at $\lz$, we must know
the first derivative, or the spectral index of the scalar spectrum at
$\lambda=\lz$:
\be
\frac{1}{A_S(\lambda)}\left.\frac{dA_S(\lambda)}{d\lambda}\right|_{\lambda=\lz}
= \left.\frac{1-n}{2\lambda}\right|_{\lambda=\lz}=\frac{1-n_0}{2\lambda_0}.
\ee
Using Eqs.\ (\ref{eq:VB}) and (\ref{eq:VZ}) we can find the first derivative
of the potential:
\be
\vzp \equiv \left. \frac{dV(\phi)}{d\phi}\right|_{\lambda=\lz} = \pm
\frac{16\pi^3}{\sqrt{2}\kappa^3} \frac{A_G^3(\lz)}{A_S(\lz)} \left[ 7-n_0 -
(5-n_0) \frac{A_G^2(\lz)}{A_S^2(\lz)} \right] .
\ee
This expression also simplifies in the $A_G(\lz) \ll A_S(\lz)$ limit:
\be
\vzp = \pm \frac{16\pi^3}{\sqrt{2}\kappa^3} \frac{A_G^3(\lz)}{A_S(\lz)}
(7-n_0) \left[ 1  + {\cal O} \left( \frac{A_G^2(\lz)}{A_S^2(\lz)} \right)
\right] \, .
\ee

Repeated differentiation of this expression with respect to $\phi$ enables one
to derive the higher derivatives. In principle, the potential  can then be
expanded as a Taylor series about the point $\phi_0$. The full expression for
the second derivative is
\bea
V''(\pz) & = & \frac{8\pi^3}{\kappa^2} \frac{A_G^2(\lz)}{A_S^2(\lz)}
\left\{ 3A_G^2(\lz)\left[ 7-n_0 -(5-n_0)\frac{A_G^2(\lz)}{A_S^2(\lz)}\right]
\right.  \nonumber \\
& & - \left( \frac{1-n_0}{2} \right)  \left[ 7-n_0 -(5-n_0)
\frac{A_G^2(\lz)}{A_S^2(\lz)} \right] [A_S^2(\lz) - A_G^2(\lz)]  \nonumber \\
& & + [A_S^2(\lz) - A_G^2(\lz) ] \left[
\left( -1+ \frac{A_G^2(\lz)}{A_S^2(\lz)}
\right) \lz n_0' \right. \nonumber \\
& &  -2(5-n_0) \frac{A_G^4(\lz)}{A_S^2(\lz)} \frac{1}{A_S^2(\lz) -
A_G^2(\lz)} \nonumber \\
& &\left.\left.+(5-n_0)(1-n_0) \frac{A_G^2(\lz)}{A_S^2(\lz)} \right] \right\} ,
\eea
where $n_0'\equiv dn_0/d\lz$. If one makes the approximation that
$A_G(\lz) \ll A_S(\lz)$, it follows that this expression simplifies
considerably:
\bea
\label{2DERSIM}
V''(\pz) & = & \frac{4\pi^3}{\kappa^2} \frac{A_G^2(\lz)}{A_S^2(\lz)}
\left[ 4(n_0 -4)^2 A_G^2(\lz) - (1-n_0)(7-n_0)A_S^2(\lz) \right] \nonumber \\
& &\times\left\{ 1 + {\cal O} \left( \frac{A_G^2(\lz)}{A_S^2(\lz)}\right) +
{\cal O}\left( \left. \frac{d n_0}{ d\ln \lambda}\right|_{\lambda=\lz} \right)
\right\}\, .
\eea
Note that $(1-n_0)$ is in principle of the same order as $A_G^2/A_S^2$.

It is hoped  that observations will soon be of a sufficient standard to
measure  $A_G(\lz)$, $A_S(\lz)$ and $n_0$ at a particular point. One may then
be able to establish whether the potential is convex or concave with the use
of Eq.\ (\ref{2DERSIM}). The potential is convex in any model where the
scalar spectrum decreases with increasing wavelength.  Note though that this
is only sufficient, not necessary. Indeed, most popular models such as
polynomial and exponential potentials are convex yet still feature a
spectrum increasing with increasing wavelength. In models where the tensor
contribution is negligible, the only important parameter is the sign of
$1-n_0$, since $n_0>7$ is already ruled out by observation

Another quantity of interest that may soon be determined observationally is
the dimensionful parameter $V(\pz)/|V'(\pz)|$. This  is determined  by the
relative amplitudes of the scalar and tensor fluctuations at a given scale via
Eq.\ (\ref{extra}).  In this sense, such a quantity yields information
regarding a mass scale at which these processes are occurring during
inflation. In the case of polynomial potentials it uniquely determines $\pz$.
For exponential and hyperbolic examples, however, it measures the steepness
of the potentials as given by $\bar{\phi}$.

Although the value of $\phi_0$ is undetermined because of the inherent freedom
to shift $\phi$ by a constant, some information of the range of $\phi$ covered
by observations of the spectra on scales between $\lz$ and $\lambda_1$ can be
recovered.  We can start with Eq.\ (\ref{eq14}), the reconstruction equation
for $\phi(\lambda)$ and find
\be
\phi_1-\phi_0 =  \pm \frac{\sqrt{2}}{\kappa } \int^{\lambda_1}_{\lambda_0}
\frac{d\lambda'}{\lambda'}
  \frac{A_S(\lambda')A_G(\lambda')}{A_S^2(\lambda')-A_G^2(\lambda')},
\ee
where $\phi_1\equiv \phi(\lambda_1)$.  Then, we can then use a simple
trapezoidal integration rule to find $\phi_1-\phi_0$ in terms of the spectra at
$\lambda_0$ and $\lambda_1$.

\vspace{48pt}
\thesection{\centerline{\bf VI. DETERMINING THE PRIMEVAL SPECTRUM}}
\setcounter{section}{6}
\setcounter{equation}{0}
\vspace{18pt}

Conventionally, one chooses a particular theory, assesses the spectrum it
predicts and attempts a comparison between its predictions and the observed
Universe. For our purposes here, one must be more ambitious and execute this
procedure in reverse order, proceeding from the observations to the primeval
spectrum, and thence to the underlying inflationary theory. As well as
covering the current observational position, we intend to survey the
possibilities inherent in future experiments, both proposed and conjectural,
in determining the primeval spectrum.\footnote{For an extensive review of
large-scale structure studies, see the papers of Efstathiou \re{E90} and
Liddle and Lyth \re{LL2}.} In keeping with our inflationary motivation, we
assume throughout that we have a universe of critical density.

The range of scales of interest stretches from the present horizon scale,
$6000 h^{-1}$ Mpc, down to about $1 h^{-1}$ Mpc, the scale which contains
roughly enough matter to form a typical galaxy.\footnote{The present density of
the Universe is $\rho_0 = 3 H_0^2/8\pi G \simeq 2.8 h^{-1} \times 10^{11}
\msun (h^{-1} {\rm Mpc})^{-3}$, where $\msun$ is the solar mass.}   On the
microwave sky, an angle of $\theta$ (for small enough $\theta$) samples linear
scales of $100 h^{-1} (\theta/1^\circ) {\rm Mpc}$.\footnote{The surface of
last scattering is located some $200h^{-1}$ Mpc inside the horizon distance.}
For purposes of discussion, it is convenient to split this range into three
separate regions.

\begin{itemize}
\item {\bf Large scales: $6000h^{-1}$ Mpc $\longrightarrow \; \; \sim 200
h^{-1}$ Mpc:}\\
These scales entered the horizon after the decoupling of the microwave
background. Except in models with peculiar matter contents, perturbations on
these scales have not been affected by any physical processes, and the
spectrum retains its original form. At present the perturbations are still
very small, growing in the linear regime without mode coupling. Here, we are
still seeing the primeval spectrum.

\item {\bf Intermediate scales: $\sim 200 h^{-1}$ Mpc $\longrightarrow 8
h^{-1}$ Mpc:}\\ These scales remain in the linear regime, and their
gravitational growth is easily calculable. However, they have been seriously
influenced by the matter content of the Universe, in a way normally specified
by a {\em transfer function}, which measures the decrease in the density
contrast relative to the value it would have had if the primeval spectrum had
been unaffected. Even in CDM models, where the only effect is the
suppression of growth due to the Universe not being completely matter
dominated at the time of horizon entry, this effect is at the $25\%$ level at
$200 h^{-1}$ Mpc. To reconstruct the primeval spectrum on these scales, it is
thus essential to have a strong understanding of the matter content of the
Universe, including dark matter, and of its influence on the growth of density
perturbations.

\item {\bf Small scales: $8 h^{-1}$ Mpc $\longrightarrow 1 h^{-1}$ Mpc:}\\ On
these scales the density contrast has reached the nonlinear regime, coupling
together modes at different wavenumbers, and it is no longer easy to calculate
the evolution of the density contrast. This can be attempted either by an
approximation scheme such as the Zel'dovich approximation
[\ref{E90},\ref{ZELD}], or more practically via $N$--body simulations
\re{NBOD}.  Further, hydrodynamic effects associated with the nonlinear
behavior can come into play, giving rise to an extremely complex problem with
important non-gravitational effects. Again, the transfer function plays a
crucial role on these scales. In hot dark matter models, perturbations on these
scales are most likely almost completely erased by free streaming, and hence no
information can be expected to be available (far less than the
rather detailed information reconstruction would require).  In a CDM model,
enough residual perturbations may remain on these scales for useful information
to be obtained.

\end{itemize}

Let us now consider each range of scales in turn, starting with the largest
scales and working down to the smallest scales.

\vspace{18pt}

\centerline{\bf A. Large scales ($6000h^{-1}$ Mpc $\longrightarrow \; \; \sim
200 h^{-1}$ Mpc):}

\vspace{18pt}

Without doubt the most important form of observation on large scales for the
near future is large-angle microwave background anisotropies. Scales of a
couple of degrees or more fall into our definition of large scales. Such
measurements are of the purest form available---anisotropy experiments
directly measure the gravitational potential at different parts of the sky, on
scales where the spectrum retains its primeval form. Such measurements also
are of interest in that the tensor modes may contribute. Tensor modes do not
participate in structure formation and most measurements we shall discuss are
oblivious to them. Further, tensor modes inside the horizon redshift away
relative to matter, and so tensor modes also fail to participate in small-angle
microwave background anisotropies.

Nevertheless, these large-scale measurements still exhibit one crucial and
ultimately uncircumventable problem. On the largest scales, the number of
statistically independent sample measurements that can be made is small. Given
that the underlying inflationary fluctuations are stochastic, one obtains only
a limited set of realizations from the complete probability distribution
function. Such a subset may insufficiently specify the underlying
distribution, which is the quantity predicted by an inflationary model, for
our purposes. This effect, which has come to be known as the {\em cosmic
variance}, is an important matter of principle, being a source of uncertainty
which remains even if perfectly accurate experiments could be carried out. At
any stage in the history of the Universe, it is impossible to accurately
specify the properties (most significantly the mean, which is what the
spectrum specifies assuming gaussian statistics) of the probability
distribution function pertaining to perturbations on scales close to that of
the observable Universe.

Observations other than microwave background anisotropies appear confined to
the long term future. Even such an ambitious project as the Sloan Digital Sky
Survey (SDSS) \re{SDSS} can only reach out to perhaps $500 h^{-1}$ Mpc, which
can only touch the lower end of our specified large scales. However, in order
to specify the fluctuations accurately, one needs many statistically
independent regions ($100$ seems an optimistic lower estimate) which means
that the SDSS may not specify the spectrum with sufficient accuracy above
perhaps $100 h^{-1}$ Mpc.

A much more crucial issue is that the SDSS will measure the galaxy
distribution power spectrum, not the mass distribution power spectrum that is
our inflationary prediction. In modern work it is taken almost completely for
granted that these are not the same, and it seems likely too that a bias
parameter (relating the two by a multiplicative constant) which remains
scale independent over a wide range of scales may be hopelessly unrealistic.
Consequently, converting from the galaxy power spectrum back to that of the
matter may require a detailed knowledge of the process of galaxy
formation and the environmental factors around distant galaxies. Once one
attempts to reach yet further galaxies with a long look-back time, one must
also understand something about evolutionary effects on galaxies. As we shall
discuss in the section on intermediate scales, it seems likely that peculiar
velocity data may be rather more informative than the statistics of the galaxy
distribution.

A more useful tool for large scales is microwave background anisotropies on
large angular scales. Our formalism closely follows that of Scaramella and
Vittorio \re{SV}. On large angular scales, the most convenient tool for
studying microwave background anisotropies is the expansion into spherical
harmonics
\be
\frac{\Delta T}{T} ({\bf x},\theta,\phi) = \sum_{l=2}^{\infty}
\sum_{m=-l}^{l} a_{lm}({\bf x}) \; Y^l_m(\theta,\phi) ,
\ee
where $\theta$ and $\phi$ are the usual spherical polars and ${\bf x}$ is the
observer position. With spherical harmonics defined as in \re{PFTV}, the
reality condition is
\be
a_{l,-m} = (-1)^m \, a_{l,m}^* .
\ee
In the expansion, the unobservable monopole term has been removed. The dipole
term has also been completely subtracted; the intrinsic dipole on the sky
cannot be separated from that induced by our peculiar velocity relative to the
comoving frame, though it is easy to show that for adiabatic perturbations it
will be negligible compared to it.

With gaussian statistics for the density perturbations, the coefficients
$a_{lm}({\bf x})$ are gaussian distributed stochastic random variables of
position, with zero mean and rotationally invariant variance depending only on
$l$
\be
\langle a_{lm}({\bf x}) \rangle = 0 \; \; \; ; \; \; \; \langle \left|
a_{lm}({\bf x})\right|^2\rangle \equiv \Sigma_l^2 .
\ee

It is crucial to note that a single observer such as ourselves sees a single
realization from the probability distribution for the $a_{lm}$. The observed
multipoles as measured from a single point are defined as
\be
Q_l^2 = \frac{1}{4\pi} \sum_{m=-l}^{l} \left| a_{lm} \right|^2 ,
\ee
and indeed the temperature autocorrelation function can be written in
terms of these
\be
C(\alpha) \equiv \left\langle \frac{\Delta T}{T} (\theta_1,\phi_1)
\frac{\Delta T}{T}(\theta_2,\phi_2) \right\rangle_{\alpha} =
\sum_{l=2}^{\infty} Q_l^2 P_l(\cos \alpha)  ,
\ee
where the average is over all directions on a single observer sky
separated by an angle $\alpha$, and $P_l(\cos \alpha)$ is a Legendre
polynomial. The expectation for the $Q_l^2$, averaged over all
observer positions, is just
\be
\langle Q_l^2 \rangle = \frac{1}{4\pi} (2l+1) \Sigma_l^2  .
\ee

A given model predicts values for the averaged quantities $\langle
Q_l^2 \rangle$. On large angular scales, corresponding to the lowest
harmonics, only the Sachs--Wolfe effect operates. One has two terms
corresponding to the scalar and tensor modes---we denote these
contributions by square brackets.  The scalar term is given by the
integral
\be
\Sigma_l^2[S] = \frac{8 \pi^2}{m^2} \int_0^{\infty} \frac{dk}{k} j_l^2
\left(2k/aH \right)\frac{2}{m^2}A_S^2 T^2(k) ,
\ee
where $j_l$ is a spherical Bessel function and the transfer function $T(k)$ is
normalized to one on large scales. As an example, a power-law spectrum $k
A_S^2 \propto k^n$ on scales where the transfer function is sufficiently close
to unity gives the oft-quoted
\be
\Sigma_l^2[S] = \Sigma_2^2[S] \; \frac{\Gamma\left(l+(n-1)/2\right)}
{\Gamma\left(l+(5-n)/2\right)} \, \frac{\Gamma\left((9-n)/2\right)}
{\Gamma\left((3+n)/2\right)} ,
\ee
which for a flat $n=1$ spectrum gives the simple $\Sigma_l^2[S] = 6
\Sigma_2^2[S]/l(l+1)$. However, true reconstruction requires the integral
expression.

The equivalent expression for the tensor modes is a rather complicated
multiple integral  which usually must be calculated numerically
[\ref{AW},\ref{STAR},\ref{LMM}]. Under many circumstances (Lucchin, Matarrese
and Mollerach \re{LMM} suggest $0.5 < n  < 1$ for power-law inflation) there is
a helpful approximation which is
that the ratio $\Sigma_l^2[S]/\Sigma_l^2[T]$ is independent of $l$ and given by
\be
\frac{\Sigma_l^2[S]}{\Sigma_l^2[T]} = \frac{2}{m^2}\frac{A_S^2}{A_G^2} .
\ee
For many purposes this is a perfectly adequate expression, but for true
reconstruction of the inflaton potential, one must of course use the exact
integral expression.

On the sky, one does not observe each contribution to the
multipoles separately. As uncorrelated stochastic variables, the expectations
add in quadrature to give
\be
\Sigma_l^2 = \Sigma_l^2[S] + \Sigma_l^2[T] .
\ee

For reconstruction purposes, there are two obstructions of principle. These
are
\begin{itemize}
\item Even if one could measure the $\Sigma_l^2$ exactly, the last scattering
surface being closed means one obtains only a discrete set of information---a
finite number of the $\Sigma_l$ covering some effective range of
scales.\footnote{The $l$-th multipole is often taken as corresponding roughly
to a scale $k_l \simeq lH_0/2 {\rm Mpc}^{-1} = l h/6000 {\rm Mpc}^{-1}$.}
There will thus be an uncountably infinite set of possible spectra which
predict exactly the same set of $\Sigma_l$.

\item One cannot measure the $\Sigma_l^2$ exactly. What one can measure is a
single realization, the $Q_l^2$. As a sum of $2l+1$ gaussian random variables,
$Q_l^2$ has a probability distribution which is a $\chi^2$ distribution
with $2l+1$ degrees of freedom, $\chi_{2l+1}^2$. The variance of this
distribution is given by
\be
{\rm Var}[Q_l^2] = \frac{2}{2l+1} \langle Q_l^2 \rangle^2,
\ee
though one should remember that the distribution is not symmetric. Each
$Q_l^2$ is a single realization from that distribution, when we really want to
know the mean. From a single observer point, there is no way of obtaining that
mean, and one can only draw statistical conclusions based on what can be
measured. Thus, a larger set of spectra which give different sets of
$\Sigma_l^2$ can still give statistically indistinguishable sets of $Q_l^2$.
The variance falls with increasing $l$ but is significant right across the
range of large scales. This is illustrated in Figure 2.
\end{itemize}

Finally, it should be mentioned that measurements of the polarization of the
CMBR on large scales may allow a separate determination of the gravitational
wave spectrum to be made \re{POLN}.  Such an effect is potentially detectable
if gravitational waves dominate the {\em COBE} result and the polarization is
of the order of $10\%$, say. If the waves only contribute $10\%$ of the {\em
COBE} signal, for example, then only $10\%$ of $10\%$ is polarized, which
significantly reduces the overall effect. Unfortunately, reconstruction of the
potential must  await a positive detection of such an effect, so we will not
discuss it further.

\vspace{18pt}

\centerline{\bf B. Intermediate scales ($\sim 200 h^{-1}$ Mpc $\longrightarrow
8 h^{-1}$ Mpc):}

\vspace{18pt}

It is on intermediate scales that determination of the primeval spectrum is
most promising, though sadly these scales only encompass about 3 $e$-foldings.
Here a range of promising observations are available, particularly towards the
small end of the range of scales. In terms of technical difficulties in
interpreting measurements, a trade-off has been made compared to large scales;
on the plus side, the cosmic variance is a much less important player as far
more independent samples are available, while on the minus side the spectrum
has been severely affected by physical processes and thus has moved a step
away from its primeval form.\footnote{In the distant future, when the horizon
size is vastly greater than at the present, there will be a range of scales
above $200 h^{-1}$ Mpc where the cosmic variance remains small and the
spectrum retains it primeval form. Such a region would be an ideal place to
carry out reconstruction, but unfortunately does not exist at the present
epoch.}

\vspace{18pt}

\centerline{1. Intermediate-scale microwave background anisotropies}

\vspace{10pt}
In the absence of reionization, the relevant angular scales are from about
$2^\circ$ down to about 5 arcminutes. (Should reionization occur, a lot of the
information on these scales could be erased or amended in difficult to
calculate ways.) Several experiments are active in this range, including the
South Pole and MAX experiments, but as none have yet published a positive
detection they are not of direct interest to us here at present. Indeed, even
with a detection many of these ground based experiments are unable to give
results with the statistical quality we would require due to the small sky
coverage which is typically involved.

Unlike the large-scale anisotropy, one cannot write down a simple expression
for the intermediate-scale anisotropies, even if it is assumed that one has
already incorporated the effect of dark matter on the growth of perturbations
via a transfer function. The reason is due to the complexity of physical
processes operating. A case in point is the expected anisotropy (specified by
the $\Sigma_l^2$, but now for larger $l$) in the CDM model
($n=1$), as calculated in detail by Bond and Efstathiou \re{BE}.

On large scales, $l^2 \Sigma_l^2$ is approximately independent of $l$ as we
have seen. Once we get into the intermediate regime, $l^2 \Sigma_l^2$ exhibits
a much more complicated form, which is dominated by a strong peak at around
$l=200$. This is induced by Thomson scattering from moving electrons at the
time of recombination. Bond and Efstathiou's calculation gives a peak height
around $6$ times as high as the extrapolated Sachs--Wolfe effect. Beyond the
first peak is a smaller subsidiary peak at $l \sim 800$.

In their calculation, Bond and Efstathiou assumed both the primeval spectrum
and the form of the dark matter. For reconstruction purposes, it seems that a
good knowledge of the form of dark matter is a pre-requisite, in order that
these processes can be calculated at all. Of course, given the number of
active and proposed dark matter search experiments, one should be optimistic
that this information will be obtained in the not too distant future. However,
even with this information, the complexity of the calculation makes it hard to
conceive of a way of inverting it, should a good experimental knowledge of the
$\Sigma_l^2$ ($l \in [30, 750]$) be obtained. Once again, it's much easier to
compare a given theory with observation than to extract a theory from
observation.

One of the interesting applications of these results might be in combination
with the large-scale measurements. The peak on intermediate scales is due only
to processes affecting the scalar modes, whereas we have pointed out that the
large-scale Sachs--Wolfe effect is a combination of scalar and tensor modes.
On large scales, one cannot immediately discover the relative normalizations
of the two contributions. However, if the dark matter is sufficiently well
understood, the height of the peak in the intermediate regime gives this
information. Should it prove that the tensors do play a significant role, then
this would be a very interesting result as it immediately excludes slow-roll
potentials for the regime corresponding to the largest scales. Should the
tensors prove negligible, then although the conclusions are less dramatic one
has an easier inversion problem on large scales as one can concentrate solely
on scalar modes.

\vspace{18pt}

\centerline{2. Galaxy clustering in the optical and infrared}

\vspace{10pt}
\noindent
{\em A. Redshift surveys in the optical.}

Over the last decade, enormous leaps have been made in our understanding of
the distribution of galaxies in the Universe from various redshift surveys.
Most prominent is doubtless the ongoing Center for Astrophysics (CfA) survey
\re{CFA}, which aims to form a complete catalogue of galaxy redshifts out to
around $100h^{-1}$ Mpc. Other surveys of optical galaxies, often trading
incompleteness for greater survey depth, are also in progress. On the horizon
is the Sloan Digital Sky Survey \re{SDSS} which aims to find the redshifts of
one million galaxies, occupying one quarter of the sky, with an overall depth
of $500 h^{-1}$ Mpc and completeness out to $100 h^{-1}$ Mpc.

The redshifts of galaxies are relatively easy (though time consuming) to
measure and interpret, and so provide one of the more observationally simple
means of determining the distribution of matter in the Universe. The main
technical problem is to correct the distribution for redshift distortions
(which gives rise to the famous ``fingers of God'' effect). However, the
distribution of galaxies, specified by the galaxy power spectrum (or
correlation function) is two steps away from telling us about the primeval
mass spectrum.
\begin{itemize}
\item We have already discussed that the primeval spectrum on intermediate
scales has been distorted by a combination of matter dynamics and amendments
to the perturbation growth rate when the Universe is not completely matter
dominated. If we know what the dark matter is, then this need not be a serious
problem.
\item Galaxies need not trace mass, and in modern cosmology it is almost
always assumed they do not. This makes the process of getting from the galaxy
power spectrum to the mass power spectrum extremely non-trivial. Models such
as biased CDM rely on the notion of a scale-independent ratio
between the two, but this too can only be an approximation to reality. In
recent work, authors have emphasised the possible influence of environmental
effects on galaxy formation (for instance, a nearby quasar might inhibit
galaxy formation \re{BW}), and indeed it has been demonstrated that only very
modest effects are required in order to profoundly affect the shapes of
measured quantities such as the galaxy angular correlation function \re{COOP}.
\end{itemize}

Despite this, attempts have been made to reconstruct the power spectrum from
various surveys. In particular, this has been done for the CfA survey
\re{CFAR}, and for the Southern Sky Redshift Survey \re{SSRS}. These
reconstructions remain very noisy, especially at both large scales (poor
sampling) and small scales (shot noise and redshift distortions), and at
present the best one could do would be to try and fit simple functional forms
such as power-laws or parametrized power spectra to them. Even then, the
constraints one would get on the slope of say a tilted CDM spectrum are very
weak indeed. However, these reconstructions go along with the usual claim that
standard CDM is excluded at high confidence due to inadequate large-scale
clustering, without providing any particular constraints on the choice of
methods of resolving this conflict.

Nevertheless, with larger sampling volumes such as those which the SDSS will
possess, one should be able to get a good determination of the {\em galaxy}
power spectrum across a reasonable range of scales, perhaps $10 h^{-1}$ to
$100 h^{-1}$ Mpc.

\vspace{12pt}
\noindent
{\em B. Redshift surveys in the infrared.}

A rival to redshifts of optical galaxies is those of infra-red galaxies, based
on galaxy positions catalogued by the Infra-Red Astronomical Satellite (IRAS)
project in the mid-eighties. The aim here is to sparse-sample these galaxies
and redshift the subset. This is being done by two groups, giving rise to the
QDOT survey \re{QDOT} and the 1.2 Jansky survey \re{1.2Jy}. Taking advantage
of the pre-existing data-base of galaxy positions has allowed these surveys to
achieve great depth with even sampling and reach some interesting conclusions.

The main obstacle to comparison with optical surveys is due to the selection
method. Infra-red galaxies are generally young, and appear to possess a
distribution notably less clustered than their optically selected
counterparts. They are thus usually attributed their own bias parameter which
differs from the optical bias. The mechanics of proceeding to the power
spectrum are basically the same as for optical galaxies.

The most interesting and relevant results here are obtained in combination
with peculiar velocity information, as discussed below.

\vspace{12pt}
\noindent
{\em C. Projected catalogues.}

As well as redshift surveys, one also has surveys which plot the positions of
galaxies on the celestial sphere. At present the most dramatic is the APM
survey \re{APM}, encompassing several million galaxies. The measured
quantity is the projected counterpart of the correlation function, the angular
correlation function usually denoted $w(\theta)$ where $\theta$ is the angular
separation. Though arguments remain as to the presence of systematics, one in
principle has accurate determinations of the galaxy angular correlation
function. The first aim is to reconstruct the full three dimensional
correlation function from this (proceeding thence to the galaxy power
spectrum). Unfortunately, present methods of carrying out this inversion
(based on inverting Limber's equation which gives $w(\theta)$ from $\xi(r)\,$)
have proven to be very unstable, and a satisfactory recovery of the full
correlation function has not been achieved.

In its preliminary galaxy identification stage, the SDSS will provide a huge
projected catalogue on which further work can be carried out.

\vspace{18pt}

\centerline{3. Peculiar velocity flows}

\vspace{10pt}
Potentially the most important measurements in large-scale structure are those
of the peculiar velocity field. Because all matter participates
gravitationally, peculiar velocities directly sample the mass spectrum, not
the galaxy spectrum. Were one to know the peculiar velocity field, this
information is therefore as close to the primeval spectrum as is microwave
background information. Indeed, in the linear regime the spectrum of the
modulus $v$ of the velocity\footnote{The spectrum is defined as ${\cal P}_v =
V (k^3/2\pi^2) \langle \left|\delta_v \right|^2 \rangle$, with $V$ being the
volume over which the Fourier components $\delta_v(k)$ are defined.} is just
given by
\be
{\cal P}_v (k) = \frac{1}{25\pi} \, \left( \frac{aH}{k} \right)^2 \,
	\frac{2}{m^2} \, A_S^2(k) T^2(k) .
\ee

Perhaps the most exciting recent development in peculiar velocity observations
is the development of the POTENT method by Bertschinger, Dekel and
collaborators \re{BERT}. Using only the assumption that the velocity can be
written as the divergence of a scalar (in gravitational instability theories
in the linear regime this is naturally associated with the peculiar
gravitational potential), they demonstrate that the radial velocity
towards/away from our galaxy (which is all that can be measured by the methods
available) can be used to reconstruct the scalar, which can then be used to
obtain the full three dimensional velocity field. This has been shown to work
very well in simulated data sets, where one mimics observations and then can
compare the reconstruction from those measurements with the original data set.
So far, the noisiness and sparseness of available real radial velocity data
has meant that attempts to reconstruct the fields in the neighborhood of our
galaxy have not yet met with great success; however, once better and more
extensive observational data are obtained one can expect this method to yield
excellent results.

At present, POTENT appears at its most powerful in combination with a
substantial redshift survey such as the IRAS/QDOT survey. As POTENT supplies
information as to the density field and the redshift survey to the galaxy
distribution, the two in combination can be used in an attempt to measure
quantities such as the bias parameter and the density parameter $\Omega_0$ of
the Universe. Reconstructions of the power spectrum have also been attempted
\re{DEKEL}. At present, the error bars (due to cosmic variance because of
small sampling volume, due to the sparseness of the data in some regions of
the sky and due to iterative instabilities) are large enough that a broad
range of spectra (including standard CDM) are compatible with the
reconstructed present-day spectrum.

With larger data sets and technical developments in the theoretical analysis
tools, POTENT (and indeed velocity data in general) appears to be a very
powerful means of investigating the present-day power spectrum. To that, one
need only add a knowledge of the dark matter to obtain the primeval spectrum
and thence to the inflaton potential. Although likely to be limited to the
range of scales specified at the lower end by the onset of the nonlinear
regime and at the upper end by the range of feasible experimental measurements
of the radial peculiar velocity, it seems that velocity data provide the most
promising means of reconstructing a segment of the inflaton potential.

\vspace{18pt}

\centerline{\bf C. Small scales ($8 h^{-1}$ Mpc $\longrightarrow 1 h^{-1}$
Mpc):}

\vspace{18pt}

It is worth saying immediately that this promises to be the least useful range
of scales. For many choices of dark matter, including the standard hot dark
matter scenario, perturbations on these scales are almost completely erased by
dark matter free-streaming to leave no information as to the primeval
spectrum. Only if the dark matter is cold does it seem likely that any useful
information can be obtained.

There are several types of measurement which can be made. Quite a bit is known
about galaxy clustering on small scales, such as the two-point galaxy
correlation function. However, the strong nonlinearity of the density
distribution on these scales erases information about the original
linear-regime structure, and the requirement of $N$-body simulations to make
theoretical predictions makes this an unpromising avenue for reconstruction
even should nature have chosen to leave significant spectral power on these
scales. There exist very small-scale (arcsecond--arcminute) microwave
background anisotropy measurements \re{OV}, though these are susceptible to a
number of line of sight effects, and further the anisotropies are suppressed
(exponentially) on short scales because the finite thickness (about $7 h^{-1}$
Mpc) of the last scattering surface comes into play.

Up to now, the most useful constraints on small scales have come from the
pairwise velocity dispersion \re{PAIR} (the dispersion of line-of-sight
velocities between galaxies). These are sensitive to the normalization of the
spectrum at small scales, though unfortunately susceptible to power feeding
down from higher scales as well. There are certainly noteworthy
constraints---for instance it is generally accepted that unbiassed standard
CDM generates excessively large dispersions. However, the calculations
required involve $N$-body simulations and because a wide range of wavelengths
contribute, obtaining knowledge of any structure in the power spectrum on these
scales is likely to prove impossible, even if the amplitude can be determined
to reasonable accuracy.

\vspace{48pt}
\thesection{\centerline{\bf VII. DISCUSSION AND CONCLUSIONS}}
\setcounter{section}{7}
\setcounter{equation}{0}

\vspace{12pt}

To date, the traditional approach in cosmology has been to take a set of
theoretical predictions  for the structure of our universe and compare them
directly with what is observed. The aim is to reduce to a minimum the space of
possible theories consistent with observations. Unfortunately such an analysis
can only deduce which theories are unsuitable and is unlikely to select
uniquely the correct one. An alternative and more ambitious program is to use
the observations to reconstruct the theory. Within the context of the
inflationary universe, for example, such an approach is justified when one
considers the prize on offer---the form of the inflaton potential. The purpose
of the present work has been to illustrate how such a reconstruction of the
potential is possible in principle.

There are two steps to any reconstruction procedure. In practice the
observational information may not be in a form which allows a direct
comparison with the theoretical predictions to be made. It is therefore
necessary to first convert the data into the quantity predicted and only then
can the second step of reconstructing the potential be completed. As was shown
in Section VI, this is especially true in the inflationary universe and
presents a number of fundamental difficulties with the procedure.

In Section III, however, we successfully completed the second step of the
process by deriving the correspondences between the tensor and scalar
fluctuation spectra and the potential. This extended the analysis of Ref.\
\re{HB} and a number of examples were presented in Section IV.
In a true reconstruction one should make no assumptions concerning the
form of the potential. In particular, the assumptions of slow-roll, which are
essentially conditions on the flatness of $V(\phi)$, should be avoided.
The formalism used places no restrictions on the inflaton field dynamics, but
does assume the slow-roll expressions for the perturbation spectra still
apply.  From a computational point of view, it follows that reconstruction is
unambiguous once the tensor spectrum is known. Unfortunately, however, it is
this quantity which is the most difficult to determine observationally. The
only observational effect of primordial gravitational waves appears to be
their influence on large-scale CMBR anisotropies. We conclude that the most
promising method of determining the tensor spectrum is to combine the
large-scale CMBR results with intermediate scale data from peculiar velocities
and CMBR anisotropies. The latter require a knowledge of the dark matter
component, but are independent of any bias in the galaxy distribution. They
determine the scalar spectrum, whereas the former depends on both the scalar
and tensor modes.  A simple subtraction therefore yields the tensor spectrum.

Eq.\ (\ref{eq:NEWSTUF}) will allow a test of the inflationary paradigm to be
made if a separate determination of the tensor spectrum on large scales can be
made. A separate determination of $A_G$ on large angular scales coupled with
{\em COBE} \re{DMR}, Tenerife \re{TEN} and the Princeton-MIT balloon
\re{MCP1991} would lead to $A_S^2$. This could then be compared with the
theoretical prediction derived from Eq.\ (\ref{eq:NEWSTUF}). If a discrepancy
was found, it would suggest that one or more of the initial assumptions---such
as the background space-time being flat; using a single, minimally coupled
scalar field or Einstein gravity---were incorrect. On the other hand, in the
absence of any discrepancy, this result could be used with a combination of
CMBR measurements around $2^\circ$, velocity and galaxy clustering data, and
compared with the theoretical predictions for different dark matter models.
This would lead to limits on the form of dark matter present in the Universe.

We note that reconstruction is still possible if the gravitational waves are
not significant, although one must then deal with the integration constant
which arises in the solution of Eq.\ (\ref{eq:NEWSTUF}) and can affect the
functional form of the potential.

Although we have been somewhat pessimistic about the near-term prospects for
reconstructing the functional form of the potential, we are optimistic
regarding the near-term possibility of obtaining some knowledge about the
potential.  To illustrate the promise of our method, let us {\em assume} that
within a few years that a combination of CMBR measurements give us some
information about the scalar and tensor amplitudes at a particular length
scale $\lambda_0$ (corresponding to an angular scale $\theta_0$).
An example is that we might in the near future have in hand the following:
\be
A_S(\lambda_0) = 1 \times 10^{-5}; \qquad A_G(\lambda_0) = 2 \times 10^{-6};
\qquad n_0 = 0.9; \qquad n_0'=0.
\ee

If we would have this information, we can follow the perturbative procedure
outlined in Section V and reconstruct information about the potential in the
vicinity of some point $\phi_0$:
\bea
V(\phi_0)   & = &      ( 2 \times 10^{16} {\rm GeV} )^4 \nonumber \\
\pm V'(\phi_0)  & = &   ( 3 \times 10^{15} {\rm GeV} )^3 \nonumber \\
V''(\phi_0) & = &      ( 5 \times 10^{13} {\rm GeV} )^2.
\eea
By taking some appropriate ratios one may find mass scales for the potential.
In this way cosmology might be first to get a ``piece of the action'' of
GUT--scale physics.

\vspace{18pt}

\thesection{\centerline{\bf ACKNOWLEDGMENTS}}

EJC and ARL are supported by the Science and Engineering Research Council
(SERC) UK. JEL is supported by an SERC postdoctoral research fellowship. EWK
and JEL are supported at Fermilab by the DOE and NASA under Grant NAGW--2381.
ARL acknowledges the use of the Starlink computer system at the University of
Sussex.  We would like to thank D.\ Lyth, P.\ J.\ Steinhardt, and M.\ S.\
Turner for helpful discussions.

\frenchspacing
\def\prl#1#2#3{{\em Phys. Rev. Lett.} {\bf #1}, #2 (#3)}
\def\prd#1#2#3{{\em Phys. Rev. D} {\bf #1}, #2 (#3)}
\def\plb#1#2#3{{\em Phys. Lett.} {\bf #1B}, #2 (#3)}
\def\npb#1#2#3{{\em Nucl. Phys.} {\bf B#1}, #2 (#3)}
\def\apj#1#2#3{{\em Astrophys. J.} {\bf #1}, #2 (#3)}
\def\apjl#1#2#3{{\em Astrophys. J. Lett.} {\bf #1}, #2 (#3)}
\vspace{1.0in}
\centerline{{\bf References}}
\begin{enumerate}
\item \label{DMR} G. F. Smoot {\em et al},  \apjl{396}{L1}{1992};
	E. L. Wright {\em et al}, \apjl{396}{L13}{1992}.
\item \label{SW} R. K. Sachs and A. M. Wolfe, \apj{147}{73}{1967};
	P. J. E. Peebles, \apjl{263}{L1}{1982}.
\item \label{GUTH} A. Guth, \prd{23}{347}{1981}.
\item \label{FOI}  D. La  and P. J. Steinhardt,  \prl{62}{376}{1989};
	E. W. Kolb, D. S. Salopek and M. S. Turner, \prd{42}{3925}{1990};
	E. W. Kolb, {\em Physica Scripta} {\bf T36}, 199 (1991);
	J. E. Lidsey, {\em Class. Quantum Grav.} {\bf  9} (1992) 149;
	J. E. Lidsey, Ph. D. Thesis, London University (1992); A. H. Guth and B.
 	Jain, \prd{45}{441}{1992}.
\item \label{NEW} A. Albrecht and P. J. Steinhardt, \prl{48}{1220}{1982};
	A. D. Linde, \plb{108}{389}{1982}.
\item \label{CHAOTIC} A. D. Linde, \plb{129}{177}{1983}.
\item \label{HB} H. M. Hodges and G. R. Blumenthal, \prd{42}{3329}{1990}.
\item \label{BARDEEN} J. M. Bardeen, \prd{22}{1882}{1980}.
\item \label{LPLB} J. E. Lidsey, \plb{273}{42}{1991}.
\item \label{POWER} F. Lucchin and S. Matarrese, \prd{32}{1316}{1985};
	\plb{164}{282}{1985}.
\item \label{ST1984} P. J. Steinhardt and M. S. Turner, \prd{29}{2162}{1984};
	E. W. Kolb and M. S. Turner, {\em The Early Universe}, (Addison-Wesley,
	New York, 1990).
\item \label{LL} A. R. Liddle and D. H. Lyth, \plb{291}{391}{1992}.
\item \label{BUNCHDAVIES} T. Bunch and P. C. W. Davies, {\em Proc. Roy.
	Soc. London} {\bf A360}, 117 (1978).
\item \label{AW} L. F. Abbott and M. B. Wise, \npb{244}{541}{1984}.
\item \label{STAR} A. A. Starobinsky, {\em Sov. Astron. Lett.} {\bf 11},
	133 (1985).
\item \label{BGKZ} V. Belinsky, L. Grischuk, I. Khalatanikov and Ya. B.
	Zel'dovich, \plb{155}{232}{1985}.
\item \label{CL1993} B. J. Carr and J. E. Lidsey, ``Primordial Black Holes
	and Generalized Constraints on Chaotic Inflation,'' QMW preprint
	(1993).
\item \label{TENSORS} L. M. Krauss and M. White, \prl{69}{869}{1992};
	J. E. Lidsey and P. Coles, {\em Mon. Not. Roy. astr. Soc.} {\bf 258},
	57P (1992); D. S . Salopek, \prl{69}{3602}{1992}; D. S. Salopek,
	in {\em Proceedings of the International School of Astrophysics
	"D. Chalogne" second course}, ed N. Sanchez (World Scientific, 1992);
	T. Souradeep and V. Sahni, {\em Mod. Phys. Lett.} {\bf A7}, 3541
	(1992); M. White, \prd{46}{4198}{1992}; R. Crittenden, J. R. Bond,
	R. L. Davis, G. Efstathiou and P. J. Steinhardt, in {\em Proceedings of
	the Texas/Pascos Symposium, Berkeley} (1992).
\item \label{DHSST} R. L. Davis, H. M. Hodges, G. F. Smoot, P. J. Steinhardt
	and M. S. Turner, \prl{69}{1856}{1992}.
\item \label{LMM} F. Lucchin, S. Matarrese, and S. Mollerach,
	\apjl{401}{49}{1992}.
\item \label{LS93} D. H. Lyth and E. D. Stewart, ``A More Accurate Analytical
	Calculation of the Spectrum of Density Perturbations Produced During
	Inflation,'' to appear {\it Phys. Lett.} {\bf B} (1993).
\item \label{LS} D. H. Lyth and E. D. Stewart, \plb{274}{168}{1992}.
\item \label{T1993} M. S. Turner, ``On the Production of Scalar and Tensor
	Perturbations in Inflationary Models,'' Fermilab preprint
	FNAL-PUB-93/026-A (1993).
\item \label{LT1993} J. E. Lidsey and R. K. Tavakol, ``On the Correspondence
	between Theory and Observation in Inflationary Cosmology,''
	Fermilab preprint FNAL-PUB- 93/034-A (1993).
\item \label{NAT} K. Freese, J. A. Frieman and A. Olinto,
	\prl{67}{3233}{1990};  F. C. Adams, J. R. Bond, K. Freese,
	J. A. Frieman and A. V. Olinto, \prd{47}{426}{1993}.
\item \label{INTINF} J. D. Barrow, \plb{235}{40}{1990}; J. D. Barrow and A. R.
	Liddle, ``Perturbation Spectra from Intermediate Inflation,'' Sussex
	preprint SUSSEX-AST 93/2-1 (1993).
\item \label{WIT} E. Witten, \npb{202}{253}{1980}.
\item \label{LL2} A. R. Liddle and D. H. Lyth, ``The Cold Dark Matter Density
	Perturbation,'' to be published, {\em Phys. Rep.} (1993).
\item \label{E90} G. Efstathiou in {\em The Physics of the Early Universe},
  	eds. A. Heavens, J. Peacock and A. Davies, SUSSP publications 1990.
\item \label{ZELD} Ya. B. Zel'dovich, {\it Astron. Astrophys.} {\bf 5}, 84
	(1970).
\item \label{NBOD} See M. Davis, G. Efstathiou, C. S. Frenk and S. D. M.
	White, {\it Nature} {\bf 356}, 489 (1992) and refs therein.
\item \label{SDSS} J. E. Gunn and G. R. Knapp, ``The Sloan Digital Sky
	Survey,'' Princeton preprint POP-488 (1992); R. G. Kron in {\em ESO
	Conference on Progress in Telescope and Instrumentation Technologies},
	ESO conference and workshop proceeding no. 42, p635, ed. M.-H. Ulrich
	(1992).
\item \label{SV} R. Scaramella and N. Vittorio, \apj{353}{372}{1990}.
\item \label{PFTV} W. H. Press, B. P. Flannery, S. A. Teukolsky and W. T.
	Vetterling, {\em Numerical Recipes}, Cambridge University Press 1986.
\item \label{POLN} A. G. Polnarev,  {\em Sov. Astron.} {\bf 29}, 607 (1985).
\item \label{BE} J. R. Bond and G. Efstathiou, {\em Mon. Not. R. astr. Soc.}
	{\bf 226}, 655 (1987).
\item \label{CFA} M. J. Geller and J. P. Huchra, {\em Science} {\bf 246},
	879 (1989); M. Ramella, M. J. Geller and J. P. Huchra,
	\apj{384}{396}{1992}.
\item \label{BW} A. Babul and S. D. M. White, {\em Mon. Not. R. astr. Soc.}
	{\bf 253}, 31p (1991).
\item \label{COOP} R. G. Bower, P. Coles, C. S. Frenk and S. D.
	M. White, \apj{405}{403}{1993}.
\item \label{CFAR} M. S. Vogeley, C. Park, M. J. Geller and J. P. Huchra,
	\apjl{391}{L5}{1992}.
\item \label{SSRS} C. Park, J. R. Gott and L. N. da Costa,
	\apjl{392}{L51}{1992}.
\item \label{QDOT} W. Saunders {\em et al}, {\em Nature} {\bf 349}, 32 (1991);
	N. Kaiser, G. Efstathiou, R. Ellis, C. Frenk, A. Lawrence, M.
	Rowan-Robinson and W. Saunders, {\em Mon. Not. R. astr. Soc.} {\bf
	252}, 1 (1991).
\item \label{1.2Jy} K. B. Fisher, M. Davis, M. A. Strauss, A. Yahil and J. P.
	Huchra, \apj{389}{188}{1992}.
\item \label{APM} S. J. Maddox, G. Efstathiou, W. J. Sutherland and J.
	Loveday, {\em Mon. Not. R. astr. Soc.} {\bf 242}, 43p (1990).
\item \label{BERT} E. Bertschinger and A. Dekel, \apjl{336}{L5}{1989}; A.
	Dekel, E. Bert\-schinger and S. M. Faber, \apj{364}{349}{1990}; E.
	Bertschinger, A. Dekel, S. M. Faber, A. Dressler and D. Burstein,
	\apj{364}{370}{1990}.
\item \label{DEKEL} A. Dekel in {\em Observational Tests of Cosmological
	Inflation}, eds T. Shanks {\it et al}, Kluwer Academic 1991.
\item \label{OV} A. C. S. Readhead, C. R. Lawrence, S. T. Myers, W. L. W.
	Sargent, H. E. Hardebeck and A. T. Moffat, \apj{346}{566}{1989}.
\item \label{PAIR} J. M. Gelb, B.-A. Gradwohl and J. A. Frieman,
	\apjl{403}{L5}{1993}.
\item \label{TEN} A. A. Watson {\it et al}, {\em Nature} {\bf 357}, 660
	(1992).
\item \label{MCP1991} S. S. Meyer, E. S. Cheng and L. A. Page,
	\apjl{371}{L1}{1991}.
\item \label{SP} T. Gaier, J. Schuster, J. Gunderson, T. Koch, M. Seiffert,
	P. Meinhold and P. Lubin, \apjl{398}{L1}{1992}.
\end{enumerate}
\newpage
\centerline{\bf Figure Captions}
\nonfrenchspacing

\noindent
{\em Figure 1:}\\
A schematic figure illustrating the main concepts behind reconstruction. For
inflation the two main steps involve converting the observations (lower half
of figure)  into the primordial scalar $(A_S)$ and tensor $(A_G)$ fluctuation
spectra and then working in reverse to reconstruct the potential $V(\phi)$.
The main observational information from the cosmic microwave background arises
through the Cosmic Background Explorer ({\em COBE}) satellite \re{DMR}, and the
Tenerife (TEN) \re{TEN} and  South Pole (SP) \re{SP} collaborations. Galaxy
surveys (APM \re{APM}, CfA \re{CFA}, IRAS [\ref{QDOT},\ref{1.2Jy}]) may
provide useful information up to $100h^{-1}$ Mpc, while the Sloan Digital Sky
Survey (SDSS) \re{SDSS} should extend to the lowest scales measured by {\em
COBE}. Peculiar velocity measurements using the POTENT (P) \re{BERT} methods
are important on intermediate scales.  The angle $\theta$ measures angular
scales on the CMBR in degrees, and length scales $\lambda$ are in units of
$h^{-1}$ Mpc. $d_H$ refers to the horizon size today and at recombination and
$d_{\rm NL} \approx 8h^{-1}$ Mpc is the scale of non-linearity. (See the text
for details). Perfect observations will only reconstruct a small portion of
the inflaton potential corresponding to between $53 \le \Delta N \le 60$
e-foldings before the end of inflation.

\vspace{2cm}

\noindent
{\em Figure 2:}\\
Multipoles up to $l=30$, roughly corresponding to the complete range of large
scales. The solid line represents the ensemble averaged $\langle Q_l^2
\rangle$ (multiplied by $l$) for a flat ($n=1$) spectrum of scalar density
perturbations with $A_G(\lambda)\ll A_S(\lambda)$ , normalized to
$\Sigma_2^2 = 1$. The three dashed lines represent different randomly chosen
realizations of this distribution. Observations can only supply one such line,
giving little clue to the ensemble average quantity that inflation supplies
the form of. For comparison, the dash-dotted line shows the result of a scalar
spectrum with $n=0.8$, again with $A_G(\lambda)\ll A_S(\lambda)$  (such a
combination would arise from an appropriate inverted harmonic oscillator
potential). Note that the normalization of this line is arbitrary (shown here
with $\Sigma_2^2 = 1$), and were it moved up it could match an observed
distribution across much of the range. More significantly, it is easy to note
that any detailed information in the spectrum such as peaks or troughs can be
swamped completely by the cosmic variance.

\end{document}